\documentclass{aa}  
\usepackage{graphicx}
\usepackage{txfonts}
\usepackage{subfigure}
\usepackage{color,soul}

\begin{document}

   \title{Stellar masses, metallicity gradients and suppressed star formation revealed in a new sample of absorption selected galaxies}

   \author{N. H. P. Rhodin\inst{1},
          L. Christensen\inst{1},
          P. M\o ller\inst{2},
          T. Zafar\inst{3},
          \and
          J. P. U. Fynbo\inst{4}
          }

   \institute{Dark Cosmology Centre, Niels Bohr Institute, Copenhagen University, Juliane Maries Vej 30, DK-2100 Copenhagen \O, Denmark\\
              \email{henrikrhodin@dark-cosmology.dk}
   \and
              European Southern Observatory, Karl-Schwarzchildstrasse 2, 85748 Garching bei M\"unchen, Germany
   \and
             Australian Astronomical Observatory, PO Box 915, North Ryde, NSW 1670, Australia
   \and
             The Cosmic Dawn Center, The Niels Bohr Institute, Copenhagen University, Juliane Maries Vej 30, 2100 Copenhagen \O, Denmark
                }

   \date{Received ; accepted }

% \abstract{}{}{}{}{} 
% 5 {} token are mandatory
 
  \abstract
   {Absorbing galaxies are selected via the detection of characteristic absorption lines which their gas-rich media imprint in the spectra of distant light-beacons. The proximity of the typically faint foreground absorbing galaxies to bright background sources makes it challenging to robustly identify these in emission, and hence to characterise their relation to the general galaxy population.}
   {We search for emission to confirm and characterise ten galaxies hosting damped, metal-rich quasar absorbers at redshift $z < 1$.}
   {We identify the absorbing galaxies by matching spectroscopic absorption -and emission redshifts and from projected separations. Combining emission-line diagnostics with existing absorption spectroscopy and photometry of quasar-fields hosting metal-rich, damped absorbers, we compare our new detections with reference samples and place them on scaling relations.}
   {We spectroscopically confirm seven galaxies harbouring damped absorbers (a $70\%$ success-rate). Our results conform to the emerging picture that neutral gas on scales of tens of kpc in galaxies is what causes the characteristic \ion{H}{i} absorption. Our key results are: (I) Absorbing galaxies with $\log _{10} [\mathrm{M}_\star ~(\mathrm{M}_\odot)] \gtrsim 10$ have star formation rates that are lower than predicted for the main sequence of star formation. (II) The distribution of impact parameter with \ion{H}{i} column density and with absorption-metallicity for absorbing galaxies at $z\sim 2-3$ extends to $z\sim 0.7$ and to lower \ion{H}{i} column densities. (III) A robust mean metallicity gradient of $\langle \Gamma \rangle = 0.022 \pm 0.001~[\mathrm{dex~kpc}^{-1}]$. (IV) By correcting absorption metallicities for $\langle \Gamma \rangle$ and imposing a truncation-radius at $12~\mathrm{kpc}$, absorbing galaxies fall on top of predicted mass-metallicity relations, with a statistically significant decrease in scatter.}
   {}


   \titlerunning{Masses, Z-gradients and SF in absorbing galaxies}
   \maketitle

%-------------------------------------------------------------------

\section{Introduction}
\label{sec:intro}

Scaling relations between galaxy observables (direct measurements and derived quantities) allow us to probe what drives galaxy evolution, and act as standard tests for simulations. Historically, scaling relations have been derived for luminosity-selected samples. In particular, such selection criteria have revealed a redshift-dependent relation between the galaxy stellar mass (M$_\star$) and the gas-phase metallicity (Z$_\text{gas}$) in the redshift range $z \sim$0.1--3.5 \citep{Tremonti2004, Savaglio2005, Erb2006, Maiolino2008}; and a relation between M$_\star$ and star formation rate (SFR) \citep{Noeske2007}. These relations were combined as projections of a fundamental redshift-invariant relation tying M$_\star$, Z$_\text{gas}$, and SFR together \citep{Mannucci2010}.

Complementing these luminosity selections, galaxies can be selected via their absorption cross-section in neutral gas when there is a chance alignment of the target with a background quasar along the line of sight. Such configurations imprint strong characteristic absorption lines in the quasar spectrum, caused by the high column density \ion{H}{i} gas in either disks, circum-galactic material (CGM) or galaxy haloes. The strongest classes of absorbers are the damped Lyman-$\alpha$ systems (DLAs) with neutral hydrogen column densities \mbox{$\log_{10} [\mathrm{N}_\mathrm{\ion{H}{i}}~(\mathrm{cm}^{-2})] \geq 20.3$} \citep{Wolfe1986}, and the sub-DLAs with column densities of \mbox{$19.0 \leq \log_{10} [\mathrm{N}_\mathrm{\ion{H}{i}}~(\mathrm{cm}^{-2})]< 20.3$} \citep{Peroux2003,Zafar2013b}. Both classes have characteristic Lorentzian damping wings associated with their Lyman-$\alpha$ profiles. Unless otherwise specified, we will refer to sub-DLAs and DLAs uniformly as \textit{damped absorbers}.

With a detectability that is independent of the host galaxy's brightness, damped absorbers are believed to sample the general galaxy population in a more representative manner, probing a larger dynamic range in stellar mass, gas-phase metallicity, star formation rate and morphology at any redshift. Indeed, this assertion was confirmed by comparing how absorption- and luminosity-selections sample the underlying luminosity function \citep{Berry2016, Krogager2017}.

This implies that a pure \ion{H}{i} selection based on damped absorbers, on average, will sample lower-mass systems often below the detection-limit in emission \citep{Fynbo1999, Moller2002, Fynbo2008, Pontzen2008, Fynbo2010, Krogager2012, Rahmati2014, Fumagalli2015}. In combination with a small projected separation to the absorption, which causes the quasar to dominate the light-throughout in a blended point spread function (PSF), this has led to a low average detection rate ($\sim 10~\%$) of confirmed counterparts \citep{Moller1998, Christensen2007, Monier2009, Fynbo2010, Fynbo2011, Meiring2011, Krogager2012, Noterdaeme2012, Peroux2012, Fynbo2013, Krogager2013, Rahmani2016}. This has prevented us from characterising the connection the majority of absorbers hold to their galaxy hosts. But presumably, sub-DLAs and DLAs on average sample different environments.

The DLA population contains the bulk of neutral gas throughout cosmic time \citep[$\Omega_\mathrm{\ion{H}{i}}^{\text{DLA}} \gtrsim 80\%$; ][]{Prochaska2005, Noterdaeme2009, Noterdaeme2012b}, and displays a weak redshift-evolution by a factor of $\sim 3$ from redshift $z\sim 5$ to the local universe \citep{Neeleman2016b}. Sub-DLAs account for the bulk of the remaining fraction in the redshift range $1.5 < z < 5.0$ \citep[$\Omega_\mathrm{\ion{H}{i}}^{\text{sub-DLA}} \sim 8-20\%$; ][]{Zafar2013a}, the Lyman limit systems (LLS, \mbox{$17 \leq \log _{10}[\mathrm{N}_\mathrm{\ion{H}{i}}~(\mathrm{cm}^{-2})] < 19.0$}) and Ly$\alpha$ forest \mbox{($\log _{10}[\mathrm{N}_\mathrm{\ion{H}{i}}~(\mathrm{cm}^2)] < 17.0$)} only contributing minor fractions to the total neutral gas content \citep{Songaila2010}.

The chemical enrichment of the DLA population evolves from $\sim 1$ percent solar metallicity at $z=5$ to $\sim 10$ percent solar in the local universe, with a $\sim 2$\,dex metallicity-spread at all redshifts \citep{Pettini1994, Ledoux2002, Prochaska2003, Rafelski2014}. This evolution and its scatter is sensitive to the underlying selection function, and can be understood as an interplay between (i) drawing from the full galaxy population at every redshift; (ii) the existence of a mass-metallicity relation at every redshift \citep{Ledoux2005, Moller2013, Christensen2014}; and (iii) a metallicity gradient within each galaxy reflecting the gradual build-up of metals in the interstellar medium (ISM) and CGM by stellar feedback and supernovae explosions. Although metallicity measurements in sub-DLAs rely on ionisation corrections \citep[see e.g.,][]{Zafar2017}, these systems are believed to be more chemically enriched than DLAs on average -in particular at low redshifts \citep{Peroux2006, Kulkarni2007, Meiring2009, Som2013, Som2015}, and there are indications that sub-DLAs arise in more massive galaxies \citep{Kulkarni2010}, which may be caused by selection biases \citep{Dessauges2009}.

The existence of a statistically significant anti-correlation between $\log _{10} [\mathrm{N}_\mathrm{\ion{H}{i}}~(\mathrm{cm}^{-2})]$ and projected separation based on photometric redshift identifications of hosts was demonstrated by \cite{Rao2011}. This is consistent with spectroscopically confirmed systems which indicate that absorbers are distributed over different impact parameters; sub-DLAs showing larger and more scattered impact parameters than DLAs \citep{Moller1998, Christensen2007, Monier2009, Fynbo2010, Fynbo2011, Meiring2011, Krogager2012, Noterdaeme2012, Peroux2012, Fynbo2013, Krogager2013, Rahmani2016}. This distribution is observed in $z \gtrsim 2$ simulations, and is attributed to the complex distribution- and flow of $\ion{H}{i}$-gas \citep{Pontzen2008, Rahmati2014}. In addition, the distribution of impact parameters correlates with absorption metallicity $([\mathrm{M}/\mathrm{H}]_\mathrm{abs})$, suggesting that DLAs probe the size of gaseous discs. This is supported by modelling DLAs as sight-lines through randomly inclined discs in high-$z$ Lyman break galaxies (LBG), which shows that the two are drawn from the same underlying distribution above $z>2$ \citep{Moller2004, Fynbo2008, Krogager2017}. 

Recent developments found a scaling relation that tie velocity-widths of metal lines in absorption to the absorber metallicity (the $\Delta V_{90}-[\mathrm{M/H}]_\mathrm{abs}$ relation). This relation is redshift-dependent \citep{Ledoux2006, Moller2013, Neeleman2013}, column-density-dependent \citep{Som2015}, and is interpreted as the absorber equivalent of the MZ-relation. Recognising that the relation can be used to target the most metal-rich absorbers has significantly increased the detection rate of absorbing galaxies in emission ($\sim 60-70~\%$), because these are associated with more massive and therefore more luminous galaxies \citep{Fynbo2010, Fynbo2011, Krogager2012, Noterdaeme2012, Peroux2012, Bouche2013, Fynbo2013, Krogager2013, Rahmani2016}. Connecting the $\Delta V_{90}$-[M/H] relation to the MZ-relation, \cite{Moller2013} predicted stellar masses in functional form $f(\tiny[\text{M/H}]_{\mathrm{abs}},~z_\mathrm{abs})$, including a free parameter $C_{[\text{M/H}]}$ to reconcile the difference between absorption- and emission-line metallicities. The relation was verified directly by comparing the stellar masses from the functional form to those derived from spectral energy distribution (SED)-fits, spanning three orders of magnitude in stellar mass \citep{Christensen2014}.

To characterise absorption selected galaxies and how they relate to the general galaxy population hinges on the low number of spectroscopically confirmed systems with complementary data in absorption and emission. Here, we attempt to rectify this issue by reporting our results from a long-slit spectroscopic follow-up of candidate hosts at redshifts $z<1$.

The paper is organised as follows: Section \ref{sec:observations} describes the sample selection, observations and archival data; Section \ref{sec:results} presents our spectroscopic measurements, extracted physical parameters and stellar masses of the absorbing galaxies. In Section \ref{sec:analysis} we combine with literature samples to investigate distributions and scaling relations. In Section \ref{sec:conclusion} we summarise our conclusions. Throughout this paper we assume a flat $\Lambda$ cold dark matter ($\Lambda$CDM) cosmology, with \mbox{H$_0$ = 70.4 kms$^{-1}$ Mpc $^{-1}$} and \mbox{$\Omega _\Lambda$ = 0.727} \citep{Komatsu2011}. 

%--------------------------------------------------------------------

\section{Observations and data-reduction}
\label{sec:observations}

\subsection{Sample selection}
\label{sec:sampleselection}
The sample represents the observed subset of targets as part of a programme to provide spectroscopic redshift-confirmation of damped absorber ($\log _{10}\text{N(\textsc{H i})} > 19.5~\text{cm}^{-2}$) counterparts at low ($z_{\text{abs}} < 1$) redshifts (programme ID 095.A-0890(A); PI: L. Christensen). We proposed ten quasar fields including 12 absorbers for long-slit spectroscopic observations with the ESO/VLT FOcal Reducer and low dispersion Spectrograph 2 \citep[FORS2;][]{Appenzeller1998}. See Section \ref{sec:FORS2data} for more details. However, only eight quasar fields containing ten absorbers were observed. Of these eight quasar fields, Q\,1436-0051 and Q\,1209+107 both contain two absorbers each. The spatial positions of the candidate-hosts of both absorbers towards \mbox{Q\,1436-0051} allowed them to be observed with a single slit placement, whilst the candidate-hosts in the Q\,1209+107 quasar field required two different slit-alignments (PA1 and PA2). In sum, we therefore report the results for nine slit alignments targeting ten absorbers.

\begin{table*}
\caption{Absorber characteristics and photometry of the candidate host galaxies identified with our FORS2 campaign.}             
\label{tab:refparam}      
\centering
\scriptsize
\begin{tabular}{*{7}{lcccccccc}}
\hline\hline
\rule{0pt}{2.5ex}Quasar field & $E_{B-V}$ $^\star$ & $z_{\text{abs}}$ & $\log_{10} [\mathrm{N}_\mathrm{\ion{H}{i}}~(\mathrm{cm}^{-2})]$ & \multicolumn{2}{c}{$[\text{M/H}]_{\text{abs}}$} & $z_{\text{phot}}$ & Filter & Magnitude & $A_{\lambda}^\star$\\ [1pt] \cline{5-6}
\rule{0pt}{2.5ex}          & [mag]            &                  &  &                      Tracer & [dex]                       &                          & & [AB]  & [mag]    \\
\hline 
\rule{0pt}{3ex}HE\,1122-1649 & 0.0369 & 0.6819 $^a$ & $20.45 \pm 0.05$ $^{a,b}$ & Fe & $-1.40 \pm 0.05$ $^{b}$ & $0.69^c$ & $U$ & $23.19 \pm 0.08$ $^c$ & $+0.05^\ddagger$\\
            &        &        &                  &         & & & $V$ & $23.01 \pm 0.06$ $^c$ & $+0.03^\ddagger$\\
            &        &        &                  &         & & & $I$ & $22.33 \pm 0.03$ $^c$ & $+0.02^\ddagger$\\
            &        &        &                  &         & & & $J$ & $22.10 \pm 0.10$ $^c$ & $+0.01^\ddagger$\\
            &        &        &                  &         & & & $H$ & $21.72 \pm 0.09$ $^c$ & $+0.01^\ddagger$\\
\rule{0pt}{3ex}Q\,0153+0009 & 0.0300 & 0.7714 $^d$ & $19.70^{+0.08}_{-0.10}$ $^{d,e}$ & Cr & $-0.52 \pm 0.20^e$ & $0.745 \pm 0.040$ $^f$ & $u'$ & $24.96 \pm 0.70$ $^g$ & $0.15$\\
           & &                   &                  & &                         & & $g'$ & $24.30 \pm 0.38$ $^f$ & $0.11$\\
           & &                   &                  & &                         & & $r'$ & $23.18 \pm 0.24$ $^f$ & $0.08$\\
           & &                   &                  & &                         & & $i'$ & $22.24 \pm 0.18$ $^f$ & $0.06$\\
           & &                   &                  & &                         & & $z'$ & $21.18 \pm 0.50$ $^f$ & $0.04$\\
           & &                   &                  & &                         & & $J$ & $21.43 \pm 0.07$ $^f$ & $0.03$\\
           & &                   &                  & &                         & & $H$ & $21.39 \pm 0.10$ $^f$ & $0.02$\\
           & &                   &                  & &                         & & $K$ & $21.01 \pm 0.07$ $^f$ & $0.01$\\
\rule{0pt}{3ex}Q\,1209+107 & 0.0217 & 0.6295 $^d$    & $20.30^{+0.18}_{-0.30}$ $^d$    & Fe & $-0.9\pm 0.40$ $^{h,\dagger}$ & $0.644 \pm 0.100$ $^f$ & $g'$  & \multicolumn{1}{r}{$<21.02$ $^f$} & - \\
               &   &                            &                                               &        & &                                   & $i'$  & $21.93 \pm 0.11$ $^f$ & - \\
               &   &                            &                                               &        & &                                   & $J$   & $21.36 \pm 0.06$ $^f$ & - \\
               &   &                            &                                               &        & &                                   & $H$   & $20.72 \pm 0.05$ $^f$ & - \\
               &   &                            &                                               &        & &                                   & $K$   & $19.89 \pm 0.03$ $^f$ & - \\
\rule{0pt}{3ex} &  & 0.3930 $^d$ & $19.46^{+0.08}_{-0.08}$ $^d$  & Zn & $0.04 \pm 0.20$ $^i$ & $0.3922\pm 0.0003$ $^{j,\star\star}$ & $u'$   & $23.21 \pm 0.18$ $^f$ & $0.11$\\
                & &              &                  &                          & & & $g'$    & $21.55 \pm 0.11$ $^f$ & $0.08$ \\
                & &              &                  &                          & & & $r'$    & $22.68 \pm 0.76$ $^f$ & $0.06$ \\
                & &              &                  &                          & & & $i'$    & $21.48 \pm 0.10$ $^f$ & $0.04$ \\
                & &              &                  &                          & & & $J$ & $22.04 \pm 0.11$ $^f$ & $0.02$ \\
                & &              &                  &                          & & & $H$ & $22.64 \pm 0.18$ $^f$ & $0.01$ \\
                & &              &                  &                          & & & $K$ & $21.83 \pm 0.13$ $^f$ & $0.01$ \\
\rule{0pt}{3ex}Q\,1323-0021      & 0.0232 & 0.7160 $^d$ & $20.40^{+0.30}_{-0.40}$ $^k$  & Zn & $0.40 ^{+0.3}_{-0.3}$ $^{k}$& $0.182 \pm 0.244$ $^f$ & $u'$   & $24.89 \pm 0.81$ $^f$ & $0.12$\\
                & &              &                                     &                      &  & & $g'$   & $22.64 \pm 0.22$ $^f$ & $0.09$\\
                & &              &                                     &                      &  & & $r'$   & $21.90 \pm 0.14$ $^f$ & $0.06$\\
                & &              &                                     &                      &  & & $i'$   & $22.03 \pm 0.23$ $^f$ & 0.05\\
                & &              &                                     &                      &  & & $m_{11200\text{\AA}}$   & $20.20 \pm 0.15$ $^k$ & $0.02$\\
                & &              &                                     &                      &  & & $m_{13500\text{\AA}}$   & $19.90 \pm 0.25$ $^k$ & $0.02$\\
                & &              &                                                             & & & & $K$   & $19.16 \pm 0.13$  $^l$ & $0.01$\\
\rule{0pt}{3ex}Q\,1436-0051      & 0.0321 & 0.7377 $^d$ & $20.08^{+0.10}_{-0.12}$ $^d$  & Zn & $-0.05 \pm 0.12$ $^{m}$          & - & $g'$           & $22.49 \pm 0.10$ $^n$ & -                 \\
                & &              &                                     &                        & & & $r'$    & $21.62 \pm 0.06$ $^n$ & - \\
                & &              &                                     &                        & & & $i'$    & $20.82 \pm 0.05$ $^n$ & - \\
                & &              &                                     &                        & & & $z'$    & $20.69 \pm 0.08$ $^n$ & - \\
                & &              &                                     &                        & & & $J$     & $20.19 \pm 0.03$ $^f$ & $0.03$\\
                & &              &                                     &                        & & & $H$     & $19.82 \pm 0.03$ $^f$ & $0.02$\\
                & &              &                                     &                        & & & $K$     & $17.54 \pm 0.05$ $^f$ & $0.01$ \\

\rule{0pt}{3ex}                &        & 0.9281 $^d$ & $18.4\pm 0.98$ $^o$                     & Zn & $-0.05\pm 0.55$ $^{o,p}$                     & - & $u'$           & $26.22 \pm 1.20$ $^f$ & $0.16$                 \\
                & &              &                                     &                        & & & $g'$     & $23.72 \pm 0.17$ $^n$ & - \\
                & &              &                                     &                        & & & $r'$     & $22.95 \pm 0.12$ $^n$ & - \\
                & &              &                                     &                        & & & $i'$     & $22.14 \pm 0.10$ $^n$ & - \\
                & &              &                                     &                        & & & $z'$     & $21.97 \pm 0.15$ $^n$ & - \\
                & &              &                                     &                        & & & $J$     & $21.07 \pm 0.06$ $^f$ & $0.03$\\
                & &              &                                     &                        & & & $H$     & $20.80 \pm 0.06$ $^f$ & $0.02$\\
                & &              &                                     &                        & & & $K$     & $18.76 \pm 0.15$ $^f$ & $0.01$\\
               
\rule{0pt}{3ex}Q\,2328+0022      & 0.0345 & 0.6519 $^d$ & $20.32^{+0.06}_{-0.07}$ $^d$  & Zn & $-0.49\pm 0.22$ $^q$                 & $0.815 \pm 0.242$ $^f$ & $u'$     & $22.60 \pm 0.78$ $^f$ & $0.17$\\
                & &              &                                     &                        & & & $g'$     & $23.23 \pm 0.14$ $^f$ & $0.13$ \\
                & &              &                                     &                         & & & $r'$    & $22.61 \pm 0.17$ $^f$ & $0.09$\\
                & &              &                                     &                        & & & $i'$     & $21.37 \pm 0.11$ $^f$ & $0.07$\\
                & &              &                                     &                        & & & $z'$     & $21.01 \pm 0.45$ $^f$ & $0.05$ \\
                & &              &                                     &                        & & & $J$   & $>20.04 \pm 0.13$ $^f$ & $0.03$\\
                & &              &                                     &                        & & & $H$   & $>19.23 \pm 0.02$  $^f$ & $0.02$\\
                & &              &                                     &                        & & & $K$   & $>19.89 \pm 0.04$  $^f$ & $0.01$\\
\rule{0pt}{3ex}Q\,2335+1501      & 0.0598 & 0.6798 $^{d,e}$      & $19.70 \pm 0.30$ $^p$        & Zn & $0.07 \pm 0.34$ $^{p}$        & - & $u'$     & $23.90 \pm 0.78$ $^g$ & $0.29$ \\
                & &              &                                     &                        & & & $g'$     & $22.76 \pm 0.14$ $^{g}$ & $0.23$\\
                & &              &                                     &                        & & & $r'$     & $21.84 \pm 0.09$ $^{g}$ & $0.16$\\
                & &              &                                     &                        & & & $i'$     & $21.33 \pm 0.09$ $^{g}$ & $0.12$\\
                & &              &                                     &                        & & & $z'$     & $21.56 \pm 0.48$ $^{g}$ & $0.09$\\

\rule{0pt}{3ex}Q\,2353-0028      & 0.0257 & 0.6044  $^d$ & $21.54^{+0.15}_{-0.15}$ $^{d,r}$ & Zn & $-0.92 \pm 0.32$ $^r$        & $0.844 \pm 0.300$ $^f$ & $u'$ & $23.44 \pm 0.59$ $^{g}$ & $0.13$\\
                &        &              &                            &                        &                &    & $g'$ & $23.35 \pm 0.23$ $^{g}$ & $0.10$\\
                &        &              &                            &                        &                  &  & $r'$ & $22.71 \pm 0.18$ $^{g}$ & $0.07$\\
                &        &              &                            &                        &                  &  & $i'$ & $21.85 \pm 0.13$ $^{g}$ & $0.05$\\
                &        &              &                            &                        &                  &  & $z'$ & $21.69 \pm 0.37$ $^{g}$ & $0.04$\\
                &        &              &                            &                        &                  & & $J$ & $20.48 \pm 0.05$ $^f$ & $0.02$\\
                &        &              &                            &                        &                  & & $H$ & $20.05 \pm 0.04$ $^f$ & $0.01$\\
                &        &              &                            &                        &                  & & $K$ & $19.27 \pm 0.02$ $^f$ & $0.01$\\
\hline
\end{tabular}

\tablefoottext{$\star$}{Galactic $E_{B-V}$ reddening and $A_\lambda$ towards the absorbing galaxy assuming the \cite{Schlafly2011} re-calibrated extinction maps.} \tablefoottext{$\star\star$}{\cite{Cristiani87} identified emission-lines $2 \sigma$ from $z_\mathrm{abs}$, 7.1 arcsec from the quasar sightline. \cite{LeBrun97} resolved the candidate into two interacting objects, the emission lines originating in one of them.}
\tablefoottext{$\dagger$}{\cite{Boisse1998}, their Table 6, report a metallicity and introduced a typical uncertainty of 0.1-0.2 dex. Due to a poorly constrained $\mathrm{N}_{\mathrm{\textsc{Hi}}}$, we assign a conservative error of 0.40 dex.}
\tablefoottext{$\ddagger$}{The magnitudes in \cite{Chen2003} were corrected for \cite{Schlegel1998} extinctions. Here we report the conversion to the \cite{Schlafly2011} maps.}
\tablebib{
\mbox{$^{(a)}$\cite{DelaVarga2000};} 
\mbox{$^{(b)}$\cite{Ledoux2002};} 
\mbox{$^{(c)}$\cite{Chen2003};} 
\mbox{$^{(d)}$\cite{RTN06};} 
\mbox{$^{(e)}$ \cite{Peroux2008};} 
\mbox{$^{(f)}$\cite{Rao2011};} 
\mbox{$^{(g)}$\citet[][SDSS DR10]{Ahn2014};} 
\mbox{$^{(h)}$ \cite{Boisse1998};} 
\mbox{$^{(i)}$ \cite{Peroux2011};}
\mbox{$^{(j)}$ \cite{Cristiani87};} 
\mbox{$^{(k)}$\cite{Moller2018};} 
\mbox{$^{(l)}$\cite{Hewett2007};} 
\mbox{$^{(m)}$ \cite{Meiring2008};} 
\mbox{$^{(n)}$ \cite{Meiring2011};} 
\mbox{$^{(o)}$ \cite{Straka2016};} 
\mbox{$^{(p)}$ \cite{Meiring2009};} 
\mbox{$^{(q)}$ \cite{Peroux2006};} 
\mbox{$^{(r)}$ \cite{Nestor2008}} 
}

\end{table*}

\begin{table*}[t!]
\caption{Observation log for the VLT/FORS2 long-slit spectroscopic observations of damped absorbing galaxies. RA and Dec refer to the coordinates of the quasar. The slit position angle (P.A.) is selected to match the candidate host galaxy seen in imaging data \citep{Rao2011}, and is defined so that N=0, E=90 degrees. All observations were taken with a slit-width of 1.31 arcsec. The tabulated airmass corresponds to the mean value calculated from the nominal header-values at start and end for each exposure. The seeing, as the FWHM of a summed profile in ten pixels along the dispersion direction, is measured in the combined spectrum of the quasar. As the seeing-measurements varied temporally and in wavelength nontrivially, we report the observed ranges. We define the systematic offset in the wavelength-solution as $\Delta \lambda _{\text{syst.}} = \lambda_{\text{calib.}} - \lambda _{\text{UVES}}$.}             
\label{tab:obslog}      
\centering
\small        
\begin{tabular}{l l l l l c c c c c}
\hline\hline
Quasar field & R.A. & Dec. & Obs. date & $\text{N}_{\text{exp.}} \times \text{t}_{\text{exp.}}$ & GRISM & Slit P.A. & Airmass & Seeing & $\Delta \lambda _{\text{syst.}}$\\
             & [J2000] & [J2000]  & [YYYY-MM-DD] & [s]                                                    & & [deg] & & [arcsec] & [\AA] \\
\hline
HE\,1122-1649                              & 11:24:42.87  & $-$17:05:17.50 & 2015-04-19      & $2\times 1800$ & 600RI & -12.90 & 1.105 & 0.86 - 0.93 & 0.33  \\
Q\,0153+0009                               & 01:53:18.19 & $+$00:09:11.44   & 2015-09-(08:09) & $4\times 1800$ & 600z  &  73.40 & 1.172 & 0.73 - 0.84 & -0.93 \\
Q\,1209+107\,PA1                            & 12:11:40.59 & $+$10:30:02.04  & 2015-04-19      & $2\times 1300$ & 600RI &  96.70 & 1.323 & 1.12 - 1.27 & -1.04 \\
Q\,1209+107\,PA2                            & 12:11:40.59 & $+$10:30:02.04  & 2015-04-19      & $2\times 1300$ & 600RI &  43.30 & 1.236 & 1.24 - 1.32 & -0.44 \\
Q\,1323-0021                               & 13:23:23.78 & $-$00:21:55.28  & 2015-04-(17:18) & $4\times 1100$ & 600RI &  42.00 & 1.158 & 1.03 - 1.17 & -0.64 \\
Q\,1436-0051                               & 14:36:45.05 & $-$00:51:50.59  & 2015-04-(13:14) & $4\times 1800$ & 600z  &  29.40 & 1.097 & 0.68 - 0.77 & -0.72 \\
Q\,2328+0022                               & 23:28:20.38 & $+$00:22:38.24   & 2015-07-11      & $2\times 800$  & 600RI &  45.00 & 1.147 & 0.78 - 0.85 & -0.38 \\
Q\,2335+1501                               & 23:35:44.19 & $+$15:01:18.37   & 2015-05-30      & $2\times 800$  & 600RI &   4.80 & 1.673 & 0.83 - 0.93 & 0.43  \\
Q\,2353-0028                               & 23:53:21.62  & $-$00:28:40.67 & 2015-07-11      & $4\times 1800$ & 600RI &  -2.30 & 1.143 & 0.90 - 0.95 & 0.35  \\
\hline
\end{tabular}

\end{table*}

We select targets with metallicities $[\mathrm{M}/\mathrm{H}]_\mathrm{abs} \gtrsim -1$ as inferred from relative element ratios measured in high resolution spectra (Table \ref{tab:refparam} and references therein), and that have a tentative galaxy counterpart based on projected separation and photometric redshift solutions consistent with $z_{\text{abs}}$ \citep{Rao2011}. Zn is minimally depleted onto dust which allows us to use it as a direct tracer-element for the absorption-metallicity (but see \cite{DeCia2018} for evidence of marginal depletion of Zn onto dust grains, in which case Zn represents a lower limit). For objects where we only have access to Fe and Cr measurements, we infer the absorption metallicity by applying a constant correction factor of 0.3 dex to account for depletion and/or $\alpha-$enhancement \citep[][our Section \ref{sec:metallicity}]{Rafelski2012}. The physical properties of the targets are summarised in Table \ref{tab:refparam}.

At $z < 1.65$, Ly$\alpha$ falls in the UV. This restricts $\mathrm{N}_\mathrm{\ion{H}{i}}$ -measurements to the systems observed with expensive and competitive space-based observatories. However, damped \ion{H}{i} absorbers are accompanied by low-ionisation metal absorption lines, including \ion{Mg}{ii} which is a tracer of DLAs \citep{RTN06}. For $z>0.11$, the \ion{Mg}{ii} $\lambda \lambda 2796, 2803~\text{\AA}$ doublet falls in the optical spectral region, where it is an efficient proxy for damped absorbers \citep[e.g. ][]{Ellison2006, Berg2017}. Our sample is therefore selected from a set of sub-DLA and DLAs with reliable HST \ion{H}{i} column-density measurements, initially identified as strong \ion{Mg}{ii} absorbers.

A large fraction of our targeted absorbers are located at redshifts $z \sim 0.7$, originating from the redshift distribution of the imaging campaigns from which they were drawn.\footnote{See for example \cite{Rao2011} where $\sim 63$\% of the systems were identified at the $3\sigma$ level in the redshift range $0.5<z<0.8$.} The distribution peaking at this value reflects the tradeoff between (i) a lower number density of LLS, sub-DLAs and DLAs towards low redshift \citep{RTN06, Zafar2013a}; and (ii) an increased limiting luminosity towards higher redshifts which prevents the identification of candidate absorbing galaxies from photometric redshift methods \citep{Rao2011}.

We target standard diagnostic nebulae lines in emission to characterise the galaxies, which allow direct comparison of the absorbing galaxies to scaling relations derived from deep imaging surveys of luminosity-selected samples at similar redshifts \citep{Savaglio2005, Maiolino2008, Karim2011, Stott2014, Whitaker2014}.

%-------------------------------------------------------------------

\subsection{VLT FORS2 data}
\label{sec:FORS2data}

Long-slit spectra were taken in nine slit-alignments.\footnote{See Section \ref{sec:sampleselection} for a detailed account of the number of fields, absorbers and slits used.} The log of observations for the different fields, including slit configuration and average seeing-conditions measured in the reduced spectra are listed in Table \ref{tab:obslog}. The observations were taken with a 1.31$''$ slit-width, the slit aligned to cover both the quasar and the candidate host with relative coordinates based on archival imaging. The observations were carried out with the 600RI and the 600z grisms, covering wavelength-ranges of $\lambda \lambda 5120-8450~\AA$ and $\lambda \lambda 7370-10700~\AA$ at a nominal spectral resolution $\mathcal{R}_\text{600RI} = 1000$ and $\mathcal{R}_\text{600z} = 1390$ assuming 1.0$''$ slit-widths, respectively \citep{Boffin2015}. For all the observations, we use a 2$\times$2 binning configuration along spatial and spectral directions.

We perform a cosmic-ray removal on the raw data frames with \verb|P3D| \citep{Sandin2010}, an adaptation of the \verb|L.A.Cosmic| algorithm \citep{vanDokkum2001}. Each frame is subsequently reduced in a standard manner by passing it through the ESO/Reflex pipeline version 5.1.4 \citep{Freudling2013}. Individual exposures are combined into 2D spectra for each object using the average pixel value and a sigma-clipping rejection. In three cases (Q\,0153+0009, Q\,1209+107 PA1, and Q\,1436-0051), the pipeline did an unsatisfactory sky-subtraction, and we post-processed the intermediate Reflex products of those observations with standard \verb|IRAF| tasks.

For each science-frame, the host-galaxy is flux-calibrated by extracting 1D-spectra of the quasar and target, scaling the quasar spectrum to its Sloan Digital Sky Survey data release 10 \cite[SDSS DR10,][]{Ahn2014} counterpart, and applying this scaling-solution to the target. For point-sources, this prevents temporally varying seeing and slit-losses from propagating into systematic errors which would be introduced in calibrating against spectrophotometric standard stars. However, it does not account for any flux variation caused by intrinsic quasar variability. Based on the the same FORS2 data for Q1323-0021, \cite{Moller2018} quantify the effect of quasar variability on the flux-calibration. In their work, they find that flux-measurements are correct to within a factor of two, with quasar variability responsible for $\sim 10\%$, and slit losses for the remaining fraction.

The SDSS catalogue does not cover HE\,1122-1649. In this field, we assume that the transmission is the same as for the other fields observed with the same instrument settings. This allows us to calculate an average sensitivity function which we use to flux calibrate the spectrum in the HE\,1122-1649 field. Based on five sensitivity functions, we calculate an RMS of $0.5 - 0.8 \times 10^{-17}~\mathrm{ergs}~\mathrm{s}^{-1}~\mathrm{cm}^{-2}$ in the wavelength region $\lambda\lambda 6000-8000~\mathrm{\AA}$. This RMS value is of the same order as integrated line-flux uncertainties (see Table \ref{tab:line-fluxes}), and accounts for weather conditions, slit-losses and quasar variability effects. We therefore confirm that the flux-calibration is robust, and verify that quasar variability, on average, is a minor effect.

To test if a systematic offset in the wavelength-solution exists and to determine velocity-dispersions, we measure central wavelengths and full-width-half-maximum (FWHM) of sky lines proximate to object emission lines in the 2D spectra. Gaussian- and Voigt profile-fitting ensured accurate central wavelengths, but systematically underestimated the FWHM measurements. The central wavelengths are measured against the Ultra-Violet and Echelle Spectrograph (UVES) telluric line catalogue \citep{Hanuschik2003}, yielding an effective offset in the wavelength-solution reported in Table \ref{tab:obslog}. With non-parametric FWHM measurements we confirm the nominal instrumental resolutions \footnote{http://www.eso.org/sci/facilities/paranal/instruments/fors/inst.html} renormalised to a 1.31 arcsec slit-width, giving effective resolutions $\mathcal{R}_\text{600RI}=763$ ($\sim 393~\text{km s}^{-1}$), and $\mathcal{R}_\text{600z}=1061$ ($\sim 283~\text{km s}^{-1}$).

%-------------------------------------------------------------------

\subsection{Archival data}
In addition to the FORS2 observations described in Sec. \ref{sec:FORS2data}, we have compiled the archival data we require in each of the targeted fields. This includes the Galactic extinction towards the lines of sight, the absorber characteristics ($z_{\mathrm{abs}}$, \mbox{$\log_{10}[\mathrm{N}_\ion{H}{i}~(\mathrm{cm}^{-2})]$}, $[\mathrm{M/H}]_{\mathrm{abs}}$) and the candidate host characteristics ($z_{\mathrm{phot}}$ and photometry).

The photometry is predominantly taken from \cite{Rao2011}, with optical images obtained at the Kitt Peak National Observatory (KPNO) with standard SDSS $u'$, $g'$, $r'$, $i'$ filters, and Near-IR (NIR) images obtained at the Mauna Kea Observatory with the NASA Infrared Telescope's NSFCAM $J$, $H$, $K$ filter set. Exceptions to this include the HE\,1122-1649 field, for which all photometry is taken with the du Pont Telescope at the Las Campanas Observatory \citep{Chen2003}; and the Q\,1323-0021 $m_{11200\text{\AA}}$ and $m_{13500\text{\AA}}$, with magnitudes measured in the collapsed 1D-spectrum \citep{Moller2018}. Where required, we apply Galactic extinction corrections to the photometric magnitudes. Here we note that the magnitudes in the HE\,1122-1649 field reported in \cite{Chen2003} were corrected for \cite{Schlegel1998} extinctions. We therefore calculate conversion factors to the re-calibrated \cite{Schlafly2011} maps and report these instead. All values and references to each entry are summarised in Table \ref{tab:refparam}.

%-------------------------------------------------------------------

\subsection{NIR photometry for Q\,1436-0051}
\label{sec:NIRphot}
The vicinity of Q\,1436-0051 has been studied extensively because it is a crowded field, hosting two absorbers in the quasar spectrum (a sub-DLA at $z_\mathrm{abs}=0.7377$  and a LLS at \mbox{$z_\mathrm{abs}=0.9281$}) and eight objects within a 30$\times$30 arcsec perimeter centred on the quasar \citep{Rao2011}. Of special interest is an object at an impact parameter of $6.4$ arcsec, where SDSS reports two source detections, classifying the combined object as a blend of an extended object with $z_\mathrm{phot} = 0.622 \pm 0.139$ and a star. \cite{Rao2011} consider the same system to be a single object (their object 3), and therefore report a combined magnitude. With optical photometry, \cite{Meiring2011} resolved the object into two extended objects (their objects 6 and 7) consistent with the redshift of the sub-DLA, likely forming an interacting pair of galaxies \citep{Meiring2011, Straka2016}.

The slit in our FORS2 observations of Q\,1436-0051 is aligned to cover this galaxy configuration. To solve for the stellar mass of each component (see Section \ref{sec:SED}), we acquired the calibrated $J$, $H$, and $K$ band images used in \cite{Rao2011} from private communication with Sandhya M. Rao, and perform aperture photometry with \verb|SExtractor| \citep{Bertin1996} version 2.19.5. We obtain isophotal flux measurements for which we require a $2\sigma$ detection threshold above the RMS noise level in the image and that a source encompasses a minimum of five pixels in the detection isophote. With these criteria in place, we resolve the two objects in the $K$-band image and ensure that we measure a representative flux-ratio of the interacting galaxies by accounting for flux in the extended wings of the flux profiles. We construct a $K$-band catalogue for object identification, and use it to extract the flux for each of the two galaxies in the $J$ and $H$ bands. The measured flux-ratios of the components in each band are combined with the total magnitudes \citep{Rao2011} to assign individual magnitudes to each of the interacting components. The resulting magnitudes are reported in Table \ref{tab:refparam}, and complement the optical photometry in the SED-fitting procedure (see Section \ref{sec:SED}).

%-------------------------------------------------------------------

\section{Results}
\label{sec:results}

\begin{figure*}[t]
    \centering
    \subfigure{\includegraphics[width=0.49\linewidth]{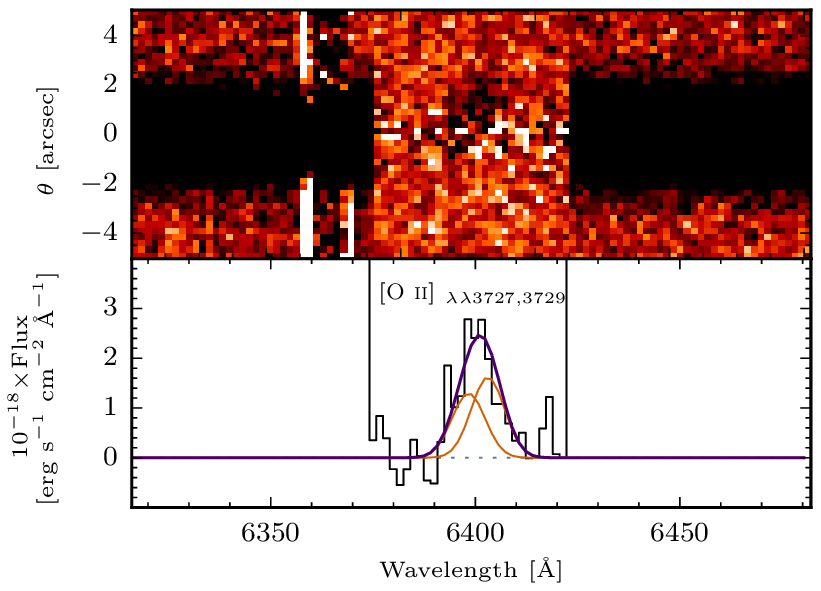}}
    \subfigure{\includegraphics[width=0.49\linewidth]{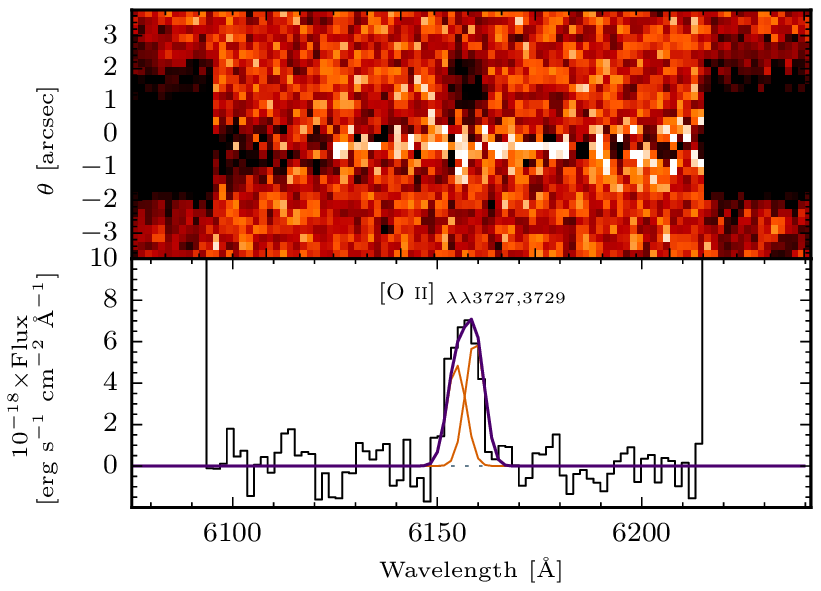}}
    \caption{Colour-inverted two-dimensional VLT/FORS2 SPSF-subtracted spectra (\textit{top panels}), and the extracted one-dimensional spectra (black) (\textit{lower panels}), centred on the [O II]$\lambda \lambda 3728,3730$ \AA~ doublet. \textit{Left:} the Q\,1323-0021 field shows excess emission from the DLA counterpart extending above the quasar-trace. \textit{Right:} Q\,2328+0022 reveals the presence of the DLA counterpart offset from the trace. In both panels, the fitted doublet-components are overplotted in orange, with the resulting emission profiles in purple.}
    \label{fig:SPSFsub}
\end{figure*}

\begin{table*}[t]
\caption{Line-flux measurements. Values for $z_{\text{em}}^{\text{SPEC}}$ refer to the heliocentric rest-frame corrected measurement, with the quoted error reflecting the uncertainty in the profile-fit. $\theta$ is the angular separation between quasar and host, measured in arcsec. $\theta$ is translated into an impact parameter ($b$) measured in kpc using the the spectroscopically determined emission-line redshift and the assumed cosmology. The fluxes have been integrated across the profile, and corrected for Galactic extinction assuming the Galactic $E_{B-V}$ values reported in Table \ref{tab:refparam}. Full-width-half-maximum (FWHM) measurements have not been corrected for the instrumental resolution.}             
\label{tab:line-fluxes}      
\centering
\small
\begin{tabular}{r | c c c | l c c c c c c c c }
\hline\hline
Quasar field \rule{0pt}{2.5ex} & \multicolumn{3}{c}{Host galaxy Identifiers} & \multicolumn{4}{c}{Emission-line Measurements}\\
 \rule{0pt}{2.5ex} & $\theta$\tablefootmark{a} & $b$\tablefootmark{b} & $z_{\text{em}}^{\text{SPEC}}$ & Transition & $\lambda _0$ & FWHM & Line-flux\\[1pt]

 \rule{0pt}{2.5ex} & ["] & [kpc] & & & [\AA] & [\AA] & [$\times 10^{-17}$ erg s$^{-1}$ cm$^{-2}$] \\
  
 \hline

HE\,1122-1649                  & 3.6\tablefootmark{c}    & 25.6    & $0.68249 \pm 0.00003$  & [\textsc{O ii}]  & 3727,3729     & $6.3 \pm 0.7$ & $17.7 \pm 2.4$   \\
                             &                         &         &                                 & H$\beta$         & 4861.39       & $6.0 \pm 0.5$ & $4.8 \pm 0.6$   \\
 
 \rule{0pt}{3ex} Q\,0153+0009  & 4.9\tablefootmark{d}    & 36.6    & $0.77085 \pm 0.00003$  & H5               & 4340.47       & $9.5 \pm 1.0$ & $3.0 \pm 0.4$    \\ 
                             &                         &         &                                 & H$\beta$         & 4861.33       & $9.5 \pm 0.6$ & $6.3 \pm 0.6$    \\ 
                             &                         &         &                                 & [\textsc{O iii}] & 5007          & $9.8 \pm 0.5$ & $7.5 \pm 0.5$ \\

 \rule{0pt}{3ex} Q\,1209+107   & 7.2\tablefootmark{d}    & 38.3    & $0.39238 \pm 0.00001$  & [\textsc{O ii}]  & 3727,3729     & $6.7 \pm 0.4$ & $41.9 \pm 1.5$   \\
                             &                         &         &                                 & H5               & 4340.47       & $5.0 \pm 0.5$ & $7.0 \pm 0.7$ \\
                             &                         &         &                                 & H$\beta$         & 4861.33       & $5.5 \pm 0.1$ & $14.8 \pm 0.4$ \\
                             &                         &         &                                 & [\textsc{O iii}] & 4958.91       & $5.6 \pm 0.1$ & $23.3 \pm 0.8$ \\
                             &                         &         &                                 & [\textsc{O iii}] & 5007          & $5.8 \pm 0.2$ & $72.1 \pm 3.0$ \\ 

 \rule{0pt}{3ex} Q\,1323-0021  & 1.4\tablefootmark{d}    & 10.2    & $0.71717 \pm 0.00008$  & [\textsc{O ii}]  & 3727,3729     & $9.6 \pm 1.7$& $3.1 \pm 0.4$  \\ 

 \rule{0pt}{3ex} Q\,1436-0051  & 6.2\tablefootmark{e}    & 45.5    & $0.73749 \pm 0.00003$  & [\textsc{O ii}]  & 3727,3729     & $10.2 \pm 0.7$\tablefootmark{g}         & $11.1 \pm 0.8$\tablefootmark{g}  \\       
                             &                         &         &                                 & H$\beta$         & 4861.33       & $7.9 \pm 1.4$ & $1.5 \pm 0.3$ \\ 
                             &                         &         &                                 & [\textsc{O iii}] & 5007          & $17.8 \pm 2.7$& $1.3 \pm 0.3$  \\

 \rule{0pt}{3ex} Q\,1436-0051  & 4.4\tablefootmark{d}    & 34.9    & $0.92886 \pm 0.00002$  & [\textsc{O iii}] & 5007          & $5.6 \pm 0.4$ & $7.0 \pm 0.7$   \\ 

 \rule{0pt}{3ex} Q\,2328+0022  & 1.7\tablefootmark{d}    & 11.9    & $0.65194 \pm 0.00006$  & [\textsc{O ii}]  & 3727,3729     & $5.5 \pm 0.7$ & $7.0 \pm 0.7$   \\

 \rule{0pt}{3ex} Q\,2335+1501  & 3.8\tablefootmark{f}    & 27.0    & $0.67989 \pm 0.00002$  & [\textsc{O ii}]  & 3727,3729     & $6.7 \pm 0.4$ & $36.3 \pm 1.4$ \\
                             &                         &         &                                 & H$\beta$         & 4861.33       & $6.5 \pm 0.2$ & $15.6 \pm 6.2$ \\ 
\hline                  
\end{tabular}

\tablefoottext{a}{Un-binned angular distance relative to the quasar.}
\tablefoottext{b}{Projected distance relative to the quasar, measured at the spectroscopic redshift.}
\tablefoottext{c}{\cite{Chen2003}}
\tablefoottext{d}{\cite{Rao2011}}
\tablefoottext{e}{This work, based on \cite{Rao2011} $K$-band image (see Section \ref{sec:NIRphot})}
\tablefoottext{f}{This work, FORS2 spectrum}
\tablefoottext{g}{This work, based on the \cite{Straka2016} spectrum (see Sections \ref{sec:galassoc}, \ref{sec:SED} and \ref{sec:starformation}).}

\end{table*}

%-------------------------------------------------------------------

\subsection{Spectral point spread function subtraction}
\label{sec:SPSF}

To scan the quasar PSF for hidden objects and recover their spectral signatures in search for the absorbing galaxies, we subtract the quasar continua in a process known as Spectral PSF (SPSF) subtraction \citep{Moller2000a, Moller2000b}. Conceptually, a model of the quasar trace in the 2D-spectrum is constructed. The model is then subtracted out to isolate line-emission from objects at small projected separations to the quasar. Variations of SPSF subtraction procedures have been used in prior studies, specifically constructed to find metal emission lines in high-redshift DLA galaxies \citep{Moller2002, Fynbo2010, Zafar2011, Fynbo2013, Krogager2013, Moller2018}. 

Here, we build a non-parametric, empirical model of the quasar PSF by averaging the observed spatial profile in two spectral windows immediately bluewards and redwards of each predicted emission-line feature for each science frame. We confirm that the quasar trace has no gradient along the dispersion direction to within sub-pixel precision. We then assume that the PSF profile does not vary with wavelength and that the quasar continuum emission can be modelled as a pure power-law across the covered wavelength range. This is used to scale the strength of the modelled quasar PSF in each wavelength, which can then be subtracted from the 2D-spectrum.

In eight out of ten targeted systems we identify the absorbing galaxy from emission lines at the expected wavelengths (see Table \ref{tab:line-fluxes}). In six of these systems the detection of emission lines is possible in the science frames without performing SPSF subtraction. This allows us to measure line-fluxes directly without the addition of noise. In passing, we note that the SPSF-subtracted spectrum of Q\,1209+107 PA2 does not reveal hidden emission lines from an object at small impact parameter. Instead, we confirm the detection of emission lines at $z=0.3922$ in the object first reported by \cite{Cristiani87} which then remains the best candidate in that field. Two additional objects are identified as the absorbing galaxies after performing SPSF subtraction (Q\,1323-0021 and Q\,2328+0022, see Figure \ref{fig:SPSFsub}). For these two systems, we vary the size and limits of the wavelength windows to see if the residual flux is real, or an artefact of the subtraction. In all realisations of the SPSF-subtraction, the residual flux is visible, which confirms that the signal is real (see Figure \ref{fig:SPSFsub}). 1D-spectra covering the recovered spectral features of the absorbing galaxies are then extracted with a simple extraction using standard IRAF tasks. By using a simple extraction, we avoid biasing the flux-measurement by the signal gradient present in the wing of the quasar PSF \citep{Moller2000a}; an effect which would result from the standard optimal extraction weighting scheme \citep{Horne1986}. We define the extraction apertures relative to the quasar trace, with aperture limits $[+0.38'':+1.26'']$ and $[+0.38'':+1.64'']$ for Q\,1323-0021 and Q\,2328+0022, respectively.

In the case of Q\,1323-0021, the spatial structure observed in the emission line feature is indicative of a velocity gradient, which makes the redshift determination  sensitive to the adopted aperture limits. We note that the extension of the signal below the quasar trace as observed in the 2D-spectrum appears to be real. However, the statistics show that this extension is driven by noise. We therefore set a lower aperture limit ensuring that we only capture the real signal of the absorbing galaxy. \cite{Moller2018} use the aperture limits $[+0.25'':+2.5'']$ above the quasar trace, attributing any further flux as quasar noise residuals from an imperfect SPSF-subtraction. We determine the redshift to be $z_{\mathrm{[\textsc{O ii}]}} = 0.71717$ (see Table \ref{tab:line-fluxes} and Table \ref{tab:results}) which is consistent with the \cite{Moller2018} measurement of \mbox{$z_{\mathrm{ref}} = 0.7170\pm 0.0006$}.

%--------------------------------------------------------------------

\subsection{Emission line measurements}
    \label{sec:emlines}

    The FORS2 wavelength coverage allows us to make line detections in [\ion{O}{ii}] $\lambda \lambda 3727,3729$ \AA, H$\beta$ $\lambda 4861 ~\mathrm{\AA}$ and \mbox{[\ion{O}{iii}] $\lambda \lambda 4959,5007$ \AA}. From here on, we use the abbreviations $[\ion{O}{ii}]$, $\mathrm{H}\beta$ and $[\ion{O}{iii}]$ to refer to these ion lines, respectively. We are able to confirm the counterparts to seven out of ten absorbers (in eight quasar fields), i.e. a conservative detection rate of $70\%$ (see Section \ref{sec:missidentification} for a discussion on an unlikely host, and Section \ref{sec:non-detections} for a discussion on the two non-detections).

We extract line fluxes by fitting Gaussian profiles to each line, with continuum placement based on a linear fit to regions bluewards and redwards of the transition free of telluric absorption and skylines. We allow the slope and normalisation of the continuum, and the centroid, linewidth and amplitude of the line profile to vary in order to retrieve the optimal fit to all lines except for the $\mathrm{[\ion{O}{ii}]}$ line doublet. At the effective resolution of the FORS2 instrument and for the target redshifts, we cannot resolve its individual components. We therefore fit two Gaussian components simultaneously, fixing the linewidths to the same value, tying the centroid-separation to the redshift, and verifying that the best fit parameters yields a line-ratio consistent with standard low-density nebulae conditions \citep{Osterbrock2006}.

Line centres are converted to velocity space, heliocentric rest-frame corrections are applied, and each velocity solution is converted to a spectroscopic redshift (see Tables \ref{tab:line-fluxes} and \ref{tab:results}). We apply Galactic extinction corrections to the line fluxes based on the $E_{B-V}$ values provided by the \cite{Schlafly2011} extinction maps. Uncertainties in the line fluxes reflect the propagated statistical errors in the profile fit.

To convert the fitted line-widths to velocity dispersions, $\sigma$, we subtract the instrumental broadening $\mathcal{R}~(\mathrm{km~s}^{-1})$ (see Sec. \ref{sec:FORS2data}) renormalised to the effective seeing, quadratically from the FWHM as
        
    \begin{equation}
        \label{eq:dispersion}
        \sigma = \sqrt{\mathrm{FWHM}^2 - \mathcal{R}^2} / (2\sqrt{2\ln{2}})~.
    \end{equation}
    
    To determine the value of each $\sigma$ and its errors, we assume that the FWHM measurements have gaussian-distributed errors. We then perturb the FWHM measurements using the measurement uncertainty in FWHM (see Table \ref{tab:line-fluxes}) to characterise the width of the distribution, centred on the measurement. Having simulated the data, we determine dispersions and errors as the median, 16th- and 84th percentile in the cumulative distribution function (CDF) of the real values below the square-root in Eq.\ref{eq:dispersion}.

%--------------------------------------------------------------------

\subsection{Galaxy associations in the Q\,1436-0051 field}
\label{sec:galassoc}

Past observations indicate that absorbing galaxies can be related to galaxy associations in dense environments \citep{Kacprzak2010, Rao2011, Christensen2014, Peroux2017}. We find a remarkable example of such associations in the field Q\,1436-0051, which is observed to host multiple galaxies that coincide with the redshifts of two damped absorbers ($\log _{10} [\mathrm{N}_\mathrm{\ion{H}{i}}~(\mathrm{cm}^{-2})] = 20.08^{+0.10}_{-0.12}$ and $18.40 \pm 0.98$, consistent with the lower limit of the definition for a sub-DLA, at redshifts $z_{\mathrm{abs}}=0.737$ and $0.928$, respectively). We confirm the presence of \textit{three} galaxies at the low-$z$, and two galaxies at the high-z, all remarkably lying within a single slit-position. We show the spatial configuration of these systems in the 2D FORS2 spectrum in the Appendix \ref{appendix_Q1436-0051}, Figure \ref{fig:Q1436-0051_groupassociations}. Based on impact-parameter and redshift information [$b$, $z_{\mathrm{em}}$] of the objects in the slit (see Table \ref{tab:appendix_Q1436-0051}), we find the most likely counterparts to be located at [45.5, 0.73749] and [34.9, 0.92886], which is consistent with Object 6 and Object 5, respectively, assuming the nomenclature of \cite{Meiring2011} and \cite{Straka2016}. We also note that the host of the $z_{\mathrm{abs}} = 0.737$ absorber is likely part of an interacting system. This issue is discussed further in Section \ref{sec:SED} and Section \ref{sec:starformation}.

%-------------------------------------------------------------------

\subsection{Unlikely host in the Q\,1209$+$107 PA2 field}
\label{sec:missidentification}

The subsequent analysis of our entire sample (see Section \ref{sec:analysis}) in relation to known scaling relations and luminosity-selected samples reveals that the candidate absorbing galaxy in the Q\,1209+107 PA2 field occupies a unique place in the MZ (Section \ref{sec:mass-metallicity}); on the SF main-sequence (\ref{sec:SFR}); and in the metallicity-gradient (Section \ref{sec:metallicity_gradient}) parameter space. In particular, it is worrying that a low-mass ($\log _{10} [\mathrm{M}_\star ~(\mathrm{M}_\odot)] \sim 8.2$) star-burst galaxy has acquired a super-solar metallicity at a distance of 38 kpc from its luminosity-centre, when the emission-line diagnostics yield a low central oxygen abundance of $[\mathrm{O/H}] \sim 8.05$.

Guided by these results and discussions, we flag the absorbing galaxy as an unlikely host. Rather, its redshift, mass, metallicity and SFR are suggestive of a smaller member in a group environment around the real host. More likely, the real host is a more massive galaxy which is hiding below the quasar PSF. We note that the SPSF-subtraction did not reveal any hidden line-flux below the quasar trace. We therefore report upper limits on the line-flux from an object under the quasar trace as the $3 \sigma$ residuals in a $\pm 5~\AA$ spectral window centred on the emission line in the SPSF-subtracted 1D-spectrum. We find consistent upper limits at a $\sim 2 \times 10^{-17} \mathrm{ergs~s}^{-1}~\mathrm{cm}^{-2}$ flux-level at the predicted position of [\ion{O}{ii}] and $\mathrm{H} \beta$ alike. 

We treat this candidate as the host in the analysis, but flag it where possible, and exclude it in calculations of the statistical properties of the sample as it most likely is not the real host of the absorber. For more details, see the relevant sections.

%--------------------------------------------------------------------

\subsection{Non-detections in the Q\,2353-0028 and Q\,1209$+$107 PA1 fields}
    \label{sec:non-detections} 
    Here we discuss the two fields (Q\,2353-0028 and Q\,1209+107\,PA1) in which the absorbing galaxy could not be identified.
    
    For Q\,2353-0028 we observe a single prominent spectral feature associated with the candidate object at $\lambda _{\mathrm{obs}} = 6565.2~\AA$. However, this does not coincide with any of the emission lines for the assumed absorber redshift. At the absorber redshift ($z_{\mathrm{abs}} = 0.6044$), this corresponds to a redshift difference of $\lvert \Delta z_{\mathrm[\textsc{O ii}]} \rvert = 0.16$, $\lvert \Delta z_{\mathrm{H}\beta} \rvert = 0.25$, and $\lvert \Delta z_{\mathrm[\textsc{O iii}]} \rvert = 0.29$, or equivalently to velocity offsets of $\lvert \Delta v_{\mathrm[\textsc{O ii}]} \rvert = 293378~\mathrm{km~s}^{-1}$, $\lvert \Delta v_{\mathrm{H}\beta} \rvert \sim 47427~\mathrm{km~s}^{-1}$, and $\lvert \Delta v_{\mathrm[\textsc{O iii}]} \rvert \sim 54778~\mathrm{km~s}^{-1}$. This leads us to refute the object as the host galaxy. We note that the emission feature is $\sim 2~\mathrm{\AA}$ redwards of the $\mathrm{H}\alpha$ rest-wavelength ($6563~\mathrm{\AA}$), which would place the galaxy at redshift $z_{\mathrm{gal}} \sim 0.0003$. However, with no additional emission lines to corroborate with, we consider the low redshift solution unlikely. We note that the quasar itself is located at redshift 0.765 \citep{RTN06}, and displays a strong [\textsc{O ii}] $\lambda \lambda 3727,3729~\mathrm{\AA}$ narrow-line doublet at the expected wavelength $\lambda_{\mathrm{[\textsc{o ii}], QSO}} = 6578~{\mathrm{\AA}}$. Interpreting the galaxy feature as [\textsc{O ii}] emission at the quasar redshift gives a velocity blue-shift $\Delta v_{\mathrm[\textsc{O ii}]} (z_{\mathrm{QSO}}) \sim 592~\mathrm{km~s}^{-1}$  between the quasar and object, which suggests that the observed emission originates from a galaxy in a group at the quasar redshift.\footnote{The object is located at an impact parameter of $\sim 5~\mathrm{arcsec}$, which at the quasar redshift for the assumed cosmology corresponds to a projected distance of $36.5~\mathrm{kpc}$.} This is also consistent with the uncertain redshift solution based on broadband photometry, $z_{\mathrm{phot}} = 0.844 \pm 0.300$ \citep{RTN06}.
    
    In the case of PA1 of the Q\,1209+107 field we do not find any spectral features of a host at the absorber redshift \mbox{($z_{\mathrm{abs}} = 0.6295$)} in the SPSF-subtracted frame, nor a stellar continuum at the position of the candidate host. The stellar-continuum of the candidate host is too faint and too proximate ($\sim 1.67$ arcsec) to be spatially resolved from the quasar PSF. This prevents us from detecting the object, also from an expected Balmer discontinuity around $\lambda 5940~\mathrm{\AA}$.

%--------------------------------------------------------------------

\subsection{Modelling the spectral energy distribution}
\label{sec:SED}
We infer stellar masses of the absorbing galaxies from modelling the spectral energy distribution (SED) with the \verb|LePhare| code \citep{Arnouts1999, Ilbert2006}, fitting it to broad-band photometric magnitudes corrected for Galactic extinction (Table \ref{tab:refparam}) using the spectroscopically determined emission-line redshifts (Table \ref{tab:results}). For matching the photometry, we use \cite[][BC03]{Bruzual2003} simple stellar population (SSP) spectral templates based on Padova 1994 stellar evolutionary tracks and a \cite{Chabrier2003} IMF. The SSPs have exponentially declining star-formation rates, parametrised with fiducial stellar population ages and e-folding time-scales in the ranges $[0.01:13.5]$ Gyrs and $[0.1:30]$ Gyrs, respectively. To model the flux in each photometric band, the filter transmission curves for each instrument used to measure the original magnitude are retrieved. 

The best-fit SED minimises the $\chi^2$-statistic across a user-defined grid of free parameters. Our grid encompasses (i) an LMC attenuation-curve \citep{Fitzpatrick2007}; (ii) a range of intrinsic reddening with $E_{B-V} \in [0.00:1.00]$ in incremental steps of 0.05 to ensure the preferred $E_{B-V}$ is associated to a $\chi^2$-minimum rather than a grid-boundary; and (iii) we allow the fits to be evaluated including- and excluding the flux contribution from nebular emission. 

\begin{figure*}
    \centering
    \subfigure{\label{fig:SED:HE1122-1649}\includegraphics[width=0.38\textwidth]{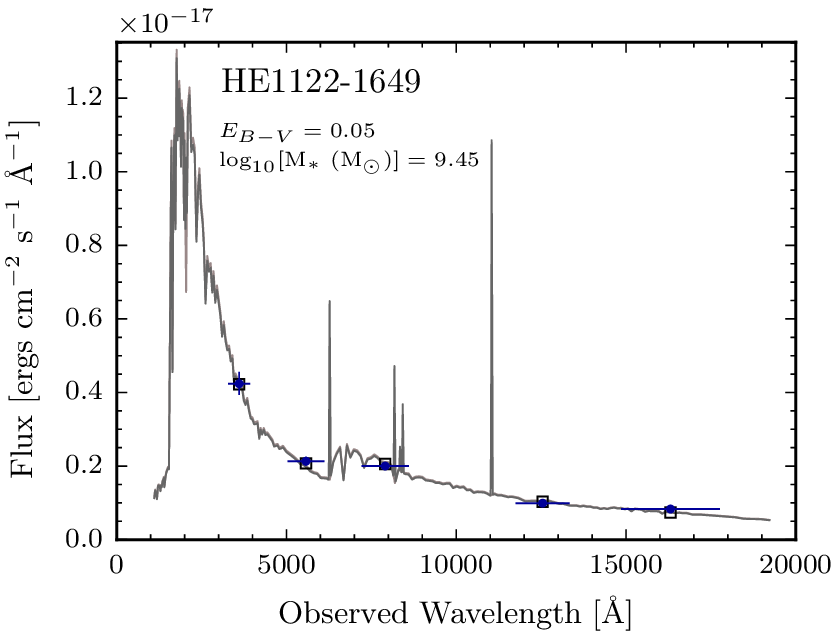}}
    \subfigure{\label{fig:SED:Q0153+0009}\includegraphics[width=0.37\textwidth]{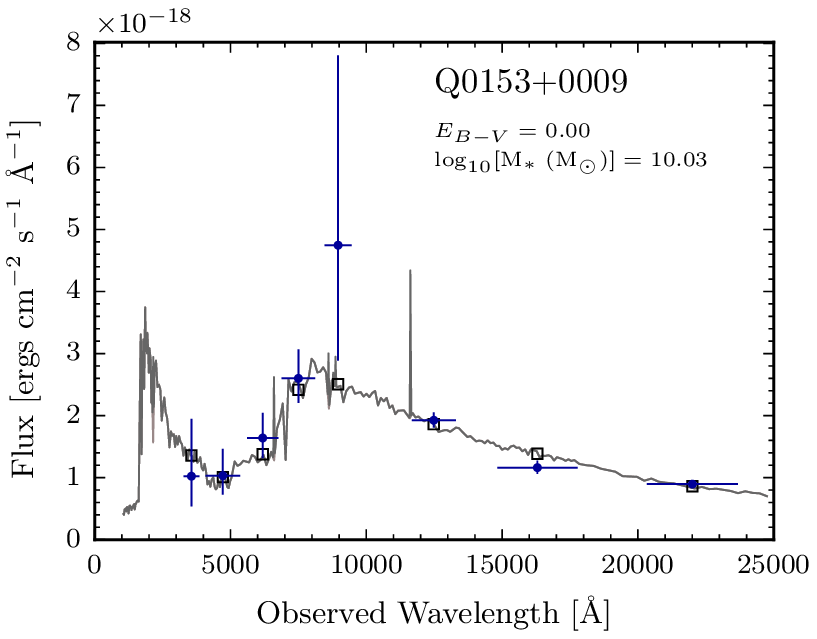}}
    \subfigure{\label{fig:SED:Q1209+107}\includegraphics[width=0.40\textwidth]{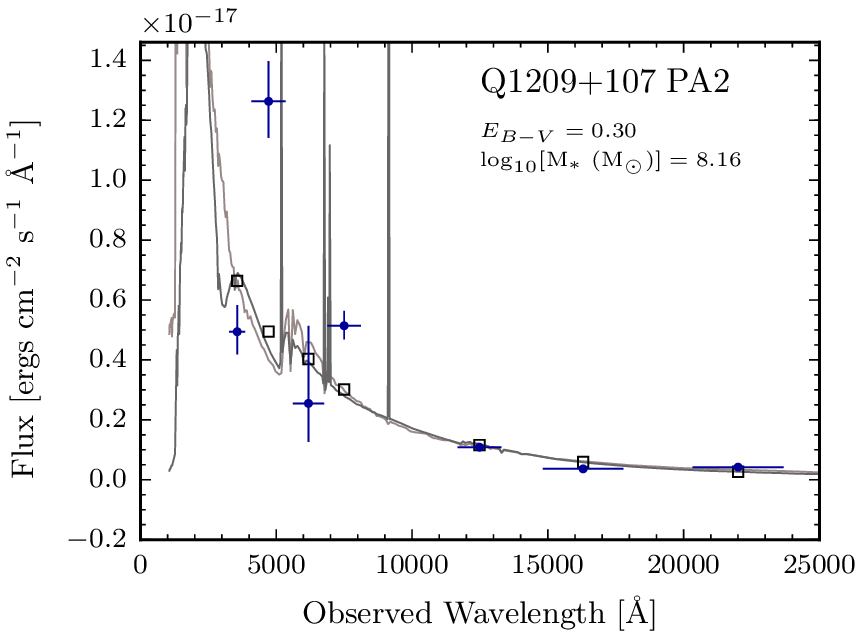}}
      \subfigure{\label{fig:SED:Q1323-0021}\includegraphics[width=0.38\textwidth]{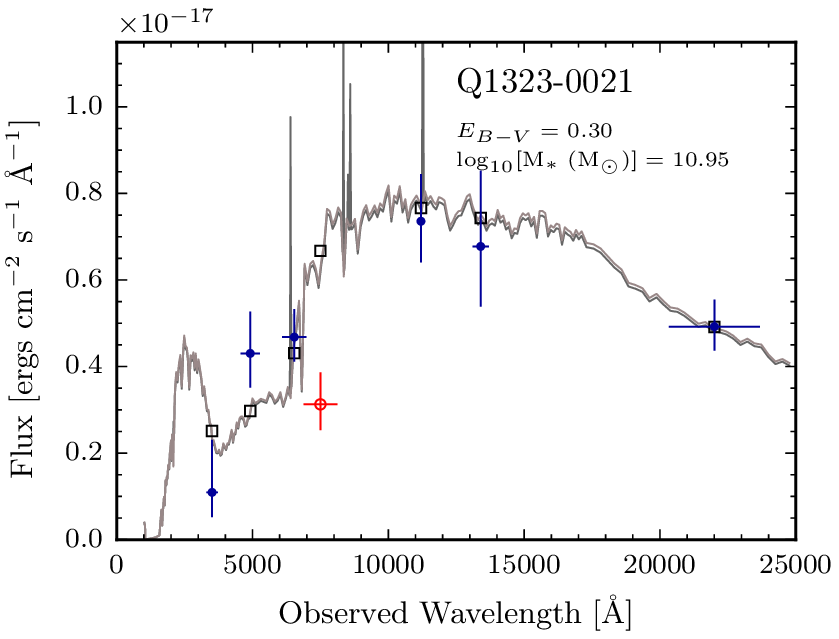}}

\subfigure{\label{fig:SED:q1436-0051_2}\includegraphics[width=0.38\textwidth]{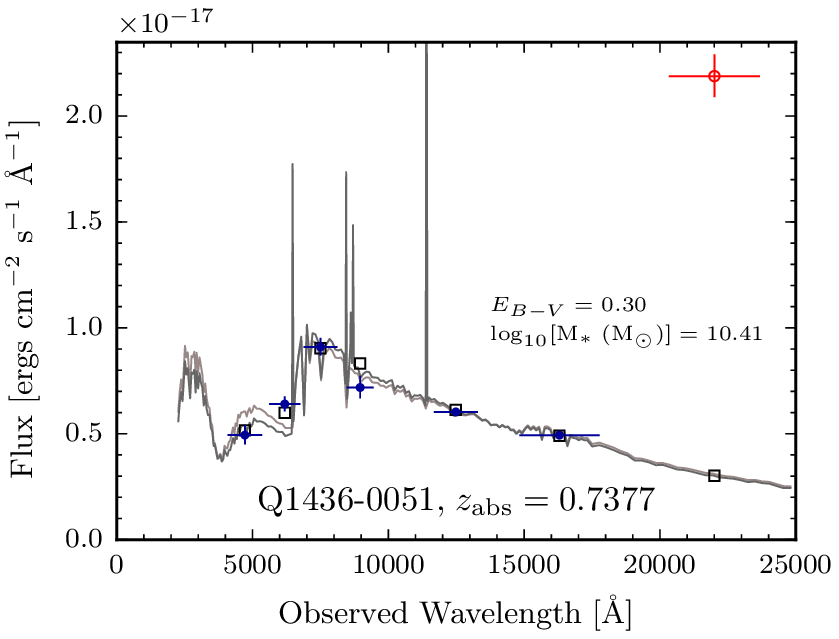}}
\subfigure{\label{fig:SED:q1436-0051_1}\includegraphics[width=0.37\textwidth]{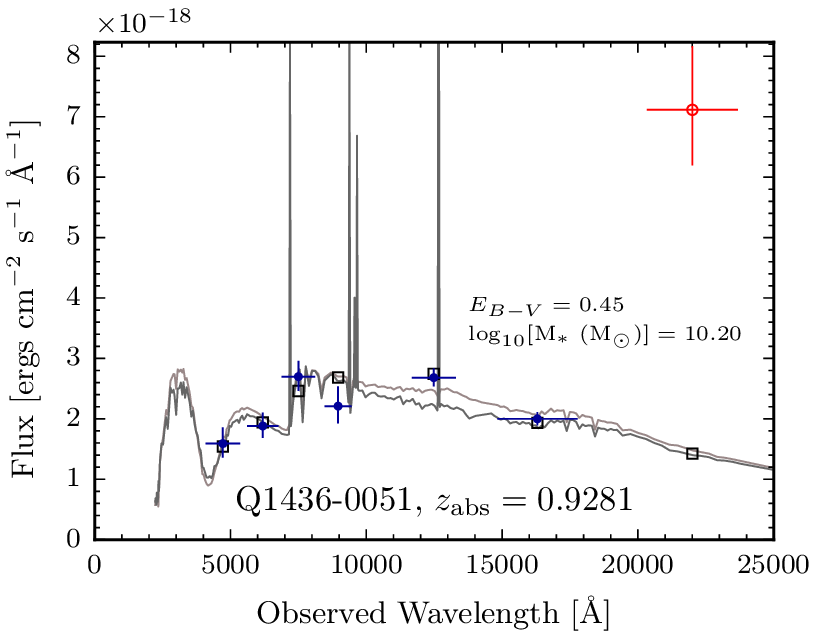}}
    \subfigure{\label{fig:SED:Q2328+0022}\includegraphics[width=0.38\textwidth]{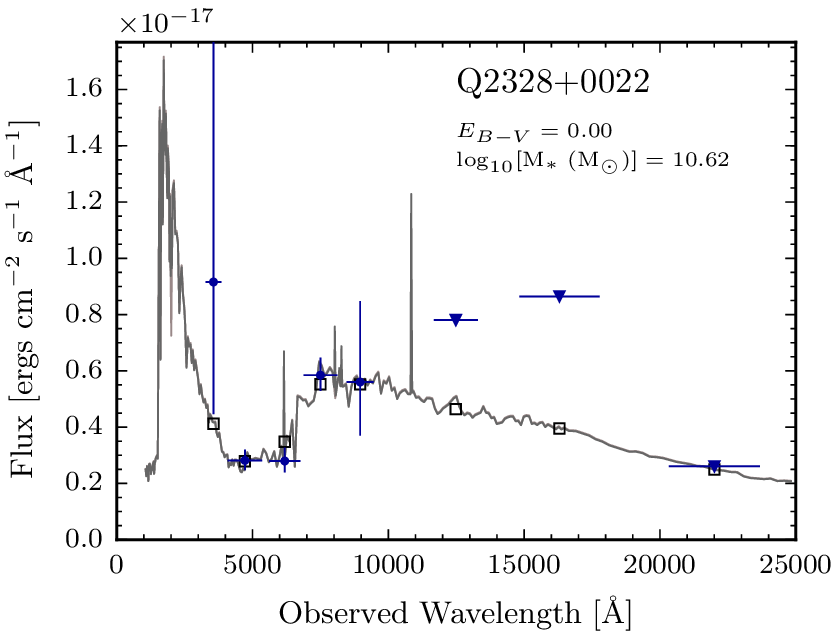}}
    \subfigure{\label{fig:SED:Q2335+1501}\includegraphics[width=0.38\textwidth]{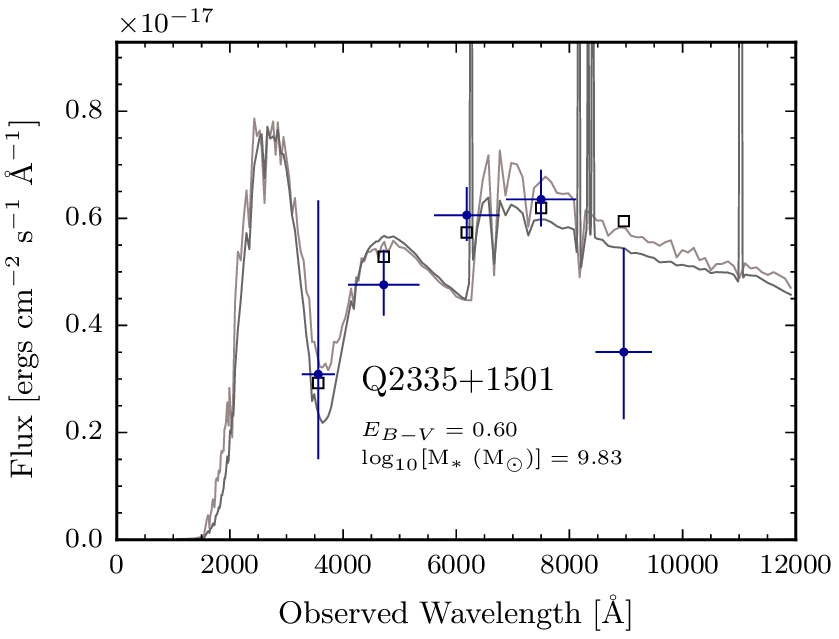}}
    \caption{The spectral energy distribution fits minimising the $\chi^2$-statistic for each of the spectroscopically identified host-galaxies. In each panel, we plot the solutions including and excluding nebular emission lines to show its impact on the derived stellar masses. Blue (red) data points refer to included (excluded) magnitudes, with vertical and horizontal error-bars indicating the uncertainty in the photometry and the FWHM of each filter, respectively. Grey squares refer to the transmission-weighted flux in each filter, calculated based on the best-fit SED solution. Upper limits are denoted by triangles. Each panel text displays the \textsc{Lephare} best fit reddening before we apply the stellar-to-nebular conversion (Equation \ref{eq:EBV_conversion}), and the resulting stellar mass.}
    \label{fig:SED_hosts}
\end{figure*}

The resulting SEDs of the absorbing galaxies are shown in Figure \ref{fig:SED_hosts}, with $\mathrm{M}_\star$ as the median value of a maximum likelihood analysis of the $\chi ^2$ distribution and $E_{B-V}$ reported in Table \ref{tab:results}. We note that for the Q\,1323-0021 field, we have priors on the nebular emission and on the $E_{B-V}$ based on detected emission lines and the dust-obscured star formation \citep{Moller2018}. For this object, we therefore prefer a model excluding the nebulae emission lines, and restrict the range of reddening to $E_{B-V} \in [0.00:0.30]$.

For Q\,1323-0021, the $i$-band magnitude was removed from the SED-fit, motivated by strong residuals from the QSO PSF-subtraction \citep{Rao2011}. In the Q\,1436-0051 field, we solve for $\mathrm{M}_\star$ for both of the interacting components (see Sections \ref{sec:NIRphot} and \ref{sec:galassoc}). In agreement with \cite{Rao2011}, we find that the $K$-band photometry is unreliable. We therefore exclude it from the SED-fitting analysis. This gives stellar masses of \mbox{$\log _{10} [\mathrm{M}_\star ~(\mathrm{M}_\odot)] = 10.41 \pm 0.10$} and \mbox{$\log _{10} [\mathrm{M}_\star ~(\mathrm{M}_\odot)] = 9.79 \pm 0.07$}, placing the absorbing galaxy in the category of a minor-merger.

\begin{table*}
\caption{Resulting parameters for our sample of absorbing galaxies. Values for $z_{\text{em}}^{\text{SPEC}}$ refer to the heliocentric rest-frame corrected measurement, with the bracket notation reflecting the uncertainty in the profile-fit. $\theta$ is the angular separation between quasar and host, measured in arcsec from known photometry. $\theta$ is translated into an impact parameter ($b$) measured in kpc using the the spectroscopically determined emission-line redshift and the assumed cosmology. $E_{B-V}^{\mathrm{nebular}}$ refers to the nebular extinctions (SED, BDEC, SFR) derived from the $\chi ^2$-minimising SED solution; the the Balmer decrement; and based on the ratio of the SFR in \mbox{$\mathrm{H}\alpha$} and \mbox{[\ion{O}{ii}]} under the assumption that they trace similar time-scales, respectively. The stellar mass and its errors are determined as the median and $1\sigma$ uncertainties from the maximum likelihood analysis of the SED $\chi ^2$-distribution. The star-formation rate estimates are corrected for intrinsic and Galactic extinction, and have been converted to a Chabrier IMF. $\sigma$ corresponds to the velocity dispersion based on the profile fits to the emission lines after resolution correction using Equation \ref{eq:dispersion}, and [O/H] to the preferred oxygen abundance derived from emission line ratios according to the M08 prescriptions. $[\mathrm{M}/\mathrm{H}]_\mathrm{em}$ refers to the metallicity in emission, inferred by applying a correction factor and truncation radius to the absorption metallicity.}
\label{tab:results}      
\centering
\small
\setlength{\tabcolsep}{1.5pt}
\begin{tabular}{r | c c c | l c c c c c c c c }
\hline\hline
\rule{0pt}{2.5ex}Quasar field & \multicolumn{3}{c}{Host-galaxy Identifiers} & \multicolumn{7}{c}{Modelled Parameter}\\
 \rule{0pt}{2.5ex} & $\theta$\tablefootmark{$\star$} & $b$ & $z_{\text{em}}^{\text{SPEC}}$ & \multicolumn{3}{c}{$E_{B-V}^{\mathrm{nebular}}$} & $\log _{10}[\mathrm{M}_\star~(\mathrm{M}_\odot)]$ & SFR$_\mathrm{[\ion{O}{II}]}$ & SFR$_{\text{H}\beta}$ & $\sigma$ & [O/H] & $[\mathrm{M}/\mathrm{H}]_\mathrm{em}$ \\[1pt] \cline{5-7}

 \rule{0pt}{2.5ex} & ["] & [kpc] & & [mag] & [mag] & [mag] & & [$\mathrm{M}_{\odot}~\mathrm{yr}^{-1}$] & [$\mathrm{M}_{\odot}~\mathrm{yr}^{-1}$] & [$\mathrm{km~s}^{-1}$] & [dex] & [dex] \\
 
 \rule{0pt}{2.5ex} & & & & SED & BDEC & SFR & & & & & \\
 
 \hline
 \rule{0pt}{3ex}HE\,1122-1649    & 3.6\tablefootmark{a}    & 25.6    & $0.68249(3)$ & $0.11$                   & -                   & $0.05^{+0.08}_{-0.04}$  & $9.45^{+0.15}_{-0.19}$    & $1.68 \pm 0.23$   & $1.55 \pm 0.19$ & $20^{+12}_{-10}$ &  $8.70^{+0.17}_{-0.20}$ \tablefootmark{$\dagger$} & $-0.84\pm0.05$ \\
 \rule{0pt}{3ex}Q\,0153+0009     & 4.9\tablefootmark{b}    & 36.6    & $0.77085(3)$ & $0.00$                   & $0.5^{+0.6}_{-0.4}$ & -  & $10.03^{+0.18}_{-0.08}$     &    -             & $16.14 \pm 1.54$ & $121^{+8}_{-8}$ & $8.75^{+0.03}_{-0.03}$ \tablefootmark{$\dagger \dagger$} & $0.04\pm 0.20$   \\
 \rule{0pt}{3ex}Q\,1209+107 PA2     & 7.2\tablefootmark{b}    & 38.3    & $0.39238(1)$ & $0.68$                   & $0.3^{+0.4}_{-0.2}$ & $\leq 0$   & $8.16^{+0.06}_{-0.06}$     & $3.20 \pm 0.12$   & $3.34 \pm 0.09$ & $49^{+22}_{-22}$ & $\sim 8.05$ \tablefootmark{$\ddagger$} & $0.30\pm 0.20$ \\
 \rule{0pt}{3ex}Q\,1323-0021     & 1.4\tablefootmark{b}    & 10.2    & $0.71717(8)$ & $0.68$  & -                   & - & $10.95^{+0.11}_{-0.14}$ \tablefootmark{$\mathsection$}   & $0.27 \pm 0.03$   & - & $134^{+44}_{-48}$  & - & $0.62\pm 0.35$    \\
 \rule{0pt}{3ex}Q\,1436-0051     & 6.2\tablefootmark{c}    & 45.5    & $0.73749(3)$ & $0.68$                   & -                   &      $\leq 0$   & $10.41^{+0.09}_{-0.08}$   & $1.02 \pm 0.07^{\star\star}$     & $0.48 \pm 0.10$       & $99^{+24}_{-26}$             & $8.82^{+0.08}_{-0.08}$ \tablefootmark{$\dagger \dagger$} & $0.21\pm 0.12$ \\
 \rule{0pt}{3ex}Q\,1436-0051     & 4.4\tablefootmark{b}    & 34.9    & $0.92886(2)$ & $1.02$                   & -                   & -           & $10.20^{+0.11}_{-0.11}$    & -     & -                  & $33^{+10}_{-12}$                &   - & $0.21\pm 0.55$ \\
 \rule{0pt}{3ex}Q\,2328+0022     & 1.7\tablefootmark{b}    & 11.9    & $0.65194(6)$ & 0.00                   & -                   & -           & $10.62^{+0.34}_{-0.36}$   & $0.47 \pm 0.05$    & -               & $56^{+23}_{-24}$ &  -    & $-0.23\pm 0.22$ \\
 \rule{0pt}{3ex}Q\,2335+1501     & 3.8\tablefootmark{d}    & 27.0    & $0.67989(2)$ & $1.36$                   & -                   & $0.11^{+0.13}_{-0.08}$        & $9.83^{+0.20}_{-0.22}$    & $4.47 \pm 0.17$  & $6.31 \pm 2.51$ & $12^{+6}_{-5}$ & $8.96^{+0.09}_{-0.26}$ \tablefootmark{$\dagger$}  & $0.33\pm 0.34 $    \\[3pt]
 \hline                  
\end{tabular}
\tablefoottext{$\star$}{Un-binned angular distance.}
\tablefoottext{$\star\star$}{Inferred from the rescaled S16 Magellan II spectrum (Sections \ref{sec:galassoc}, \ref{sec:SED} and \ref{sec:starformation}).}
\tablefoottext{$\mathsection$}{Upper limit on $E_{B-V}$ imposed from dust-content priors (Section \ref{sec:SED}).}
\tablefoottext{$\dagger$}{$\mathrm{O}_2$ diagnostic measurement.}
\tablefoottext{$\dagger \dagger$}{$\mathrm{O}_3$ diagnostic measurement.}
\tablefoottext{$\ddagger$}{$\mathrm{R}_{23}$ diagnostic measurement.}
\tablebib{$^{(a)}$ \cite{Chen2003}; $^{(b)}$ \cite{Rao2011}; $^{(c)}$ This work, based on \cite{Rao2011} $K$-band image (Section \ref{sec:NIRphot}); $^{(d)}$ This work.}

\end{table*}

%--------------------------------------------------------------------        

\subsection{Intrinsic extinction correction}
    \label{sec:EBV}   
   For three galaxies in the fields: Q\,0153+0009, Q\,1209+107 PA2 and Q\,1436-0051, we detect higher order Balmer-lines (see Table \ref{tab:line-fluxes} for individual measurements). This allows us to correct line-fluxes for intrinsic dust extinction based on the Balmer decrement (Equation \ref{eq:EBV_lrat}). The broad-band color excess is tied to the deviation in observed-to-intrinsic line-ratios as
    \begin{equation}
    \label{eq:EBV_lrat}
    E_{B-V} = \frac{2.5}{\kappa (\mathrm{H}5) - \kappa(\mathrm{H}\beta)} \log _{10} \Bigg(\frac{(\mathrm{H}\beta / \mathrm{H}5)_{\mathrm{obs}}}{(\mathrm{H}\beta / \mathrm{H}5)_0} \Bigg)~,
\end{equation}
where $\kappa$ is the value of the attenuation curve at the rest-wavelength of the labeled line-transition and $(\mathrm{H}\beta / \mathrm{H}5)_{\mathrm{obs}}$ is the observed flux ratio based on the entries in Table \ref{tab:line-fluxes}. $(\mathrm{H}\beta / \mathrm{H}5)_0$ is the intrinsic flux ratio with an adopted value of $2.137$ assuming case B recombination at $T = 10^4~\mathrm{K}$ and $n_e \sim 10^2 - 10^4~\mathrm{cm}^{-3}$ \citep{Osterbrock2006}. We adopt the \cite{Fitzpatrick2007} LMC attenuation-curve with $R_V = A_V / E_{B-V} = 3.1$, but note that extinction corrections based on Balmer lines are relatively insensitive to the choice of attenuation-curve as these behave similar redwards of the 2175\,\AA~ extinction bump.

We also attempt to use H6, H7 and H8 measurements (see Table \ref{tab:line-fluxes}) to constrain the intrinsic extinction where these lines are observed. However, the flux in the higher order lines decreases rapidly, which introduces large errors. We also attempt to model the observed line-ratios simultaneously with extinction as a free parameter. However, the large measurement errors in the higher order lines prevents us from better constraining the fit. We therefore take a conservative approach, and use the values based on the strongest Balmer decrement $(\mathrm{H}5/\mathrm{H}\beta)$ in all cases. We intentionally avoid combining our FORS2 H$\beta$ measurements with literature H$\alpha$ measurements (see for example \cite{Moller2018} for a Q1323-0021 H$\alpha$ measurement based on Sinfoni IFU data), since such a combination would introduce systematic effects.

Under the assumption that the $\mathrm{H}\alpha$ and [\textsc{O ii}] star formation rate indicators (Section \ref{sec:starformation}, eq. \ref{eq:SFR_Ha} and \ref{eq:SFR_Oii}) trace similar timescales, any discrepancy between the two can be attributed to abundance effects and dust reddening \citep{Kewley2004}. Requiring that the two be equal, the extinction then becomes
\begin{equation}
    \label{eq:EBV_SFR}
    E_{B-V} = \frac{2.5}{\kappa_{\mathrm{[\textsc{O ii}]}} - \kappa_{\mathrm{H}\alpha}} \log _{10} \Bigg(\frac{\mathrm{SFR}_{\mathrm{H\alpha}}}{\mathrm{SFR}_{\mathrm{[\textsc{O ii}]}}} \Bigg)~,
\end{equation}
where SFR$_{\mathrm{H}\alpha}$ is inferred from the measured H$\beta$ flux (see Section \ref{sec:starformation}). Where no direct spectroscopic measurements can be used to quantify the reddening, the $E_{B-V}$ can instead be inferred from the spectral energy distribution fits (see Sec. \ref{sec:SED}). The color excess of the stellar continuum has empirically been tied to the color excess derived from the nebular gas emission lines \citep{Calzetti1997, Calzetti2000}, such that
\begin{equation}
    \label{eq:EBV_conversion}
    E_{B-V}^{\mathrm{stellar}} = 0.44 \times E_{B-V}^{\mathrm{nebular}}~.
\end{equation}

The resulting $E_{B-V}$ values are reported in Table \ref{tab:results}. For completeness, we report all $E_{B-V}$ estimates. Hereafter, we distinguish them using the abbreviations BDEC (Balmer decrement), SFR (star formation rate), and SED (spectral energy distribution) to reflect the method with which the measurements were inferred. We do not detect all emission lines required to determine the $E_{B-V}$ based on both BDEC and SFR for any object. Without having to select between the two methods, we therefore proceed to apply intrinsic $E_{B-V}$ corrections evaluated from the line measurements when computing the SFRs and [O/H]s reported in Table \ref{tab:results}. SED-based $E_{B-V}$ are provided to display our best-fit SED results and for comparing measurements of $E_{B-V}$ with different techniques, but we do not correct any flux measurements with these values. The SED-based $E_{B-V}$ are poorly constrained by the lack of UV-photometry, and are only used to define a grid on which to optimise the SED fits to obtain stellar masses (see Section \ref{sec:SED}).

%--------------------------------------------------------------------

\subsection{Star formation}
    \label{sec:starformation}
We measure the unobscured star formation rates of the host galaxies with nebulae emission-line diagnostics assuming the standard $\mathrm{H}\alpha~\lambda 6563~\mathrm{\AA}$ \cite{Kennicutt1998} -and the re-calibrated $\mathrm{[\ion{O}{ii}]}~\lambda \lambda 3727,3729~\AA$ \cite{Kewley2004} relation:

\begin{equation}
    \label{eq:SFR_Ha}
    \begin{split}
    \mathrm{SFR}_{\mathrm{H}\alpha \mathrm{,10Myr}}[\mathrm{M}_\odot~\mathrm{yr}^{-1}] & = 7.9 \times 10^{-42} \mathrm{L}_{\mathrm{H}\alpha}~[\mathrm{ergs~s}^{-1}] \\
                                                                                       & = 2.3 \times 10^{-41} \mathrm{L}_{\mathrm{H}\beta}~[\mathrm{ergs~s}^{-1}]~,
    \end{split}
\end{equation}

\begin{equation}
    \label{eq:SFR_Oii}
    \mathrm{SFR}_{\mathrm{[\ion{O}{ii}]}\mathrm{,10Myr}}[\mathrm{M}_\odot~\mathrm{yr}^{-1}] = 6.58 \times 10^{-42} \mathrm{L}_{\mathrm{[\ion{O}{ii}]}}~[\mathrm{ergs~s}^{-1}]~.
\end{equation}
$\mathrm{L_i}$ refers to the dust-corrected intrinsic luminosity in SFR calibrator i. It is determined using the flux $f_i$ and the luminosity-distance $d_L$ using the python \verb|astropy.cosmology| tool \citep{Astropy2013} for the assumed cosmology, evaluated for the absorbing galaxy emission-line redshift $z_{em}$ (see Section \ref{sec:emlines} and Table \ref{tab:results}), as $\mathrm{L}_i = \mathrm{L}_{\mathrm{obs}} \times 10^{0.4 \kappa(\lambda) E_{B-V}^{\mathrm{nebular}}}$. Here, $\mathrm{L}_{\mathrm{obs}} = f_i 4 \pi d_L^2$. We note that in the absence of a dust-correction from reliable spectroscopic $E_{B-V}$ measurements (see Section \ref{sec:EBV}), the SFR can be assumed as a lower limit. The $\mathrm{H}\alpha$ luminosity is inferred from a conversion of the $\mathrm{H\beta}$ luminosity, assuming the intrinsic line ratio $\mathrm{H}\alpha / \mathrm{H}\beta = 2.86$ for standard case B recombination \citep{Osterbrock2006}.

The calibration of these relations assume solar abundances and a Salpeter initial mass function (IMF; \citealt{Salpeter1955}) with lower- and upper stellar mass cutoffs at $0.1$ and $100~\mathrm{M}_\odot$ respectively. As a Hydrogen recombination line, the $\mathrm{H}\alpha$ flux is sensitive to the incident radiation blueward of the 912 Lyman limit. This makes recombination lines direct probes of young ($\leq 10~\mathrm{Myrs}$), massive ($> 10~\mathrm{M}_\odot$) stars which dominate the ionising photon budget, and therefore tracers of the near-instantaneous SFR on time-scales of $\sim 10~\mathrm{Myrs}$. The $\mathrm{SFR}_\mathrm{[\textsc{O ii}]}$ relation is based on the $\mathrm{[\textsc{O ii}]}~\lambda \lambda 3727,3729~\AA$ collisionally excited forbidden-line doublet, empirically calibrated against $\mathrm{H}\alpha$ to give a quantitative SFR tracer on a similar time-scale \citep{Gallagher1989, Kennicutt1992, Kewley2004}. For internal consistency, we shift the values to a Chabrier IMF \citep{Chabrier2003} by applying a downward correction factor of 1.8. For objects where both SFR measurements are possible, we assign preference to SFR$_{\mathrm{H}\beta}$ as this is the most direct probe of star formation, and since $\mathrm{SFR}_\mathrm{[\textsc{O ii}]}$ is calibrated to this recombination line (see Section \ref{sec:SFR} for more information). The resulting SFRs based on both diagnostics and their associated measurement errors are summarised in Table \ref{tab:results}. 

Emission from the [\ion{O}{ii}] doublet was used by \cite{Straka2016} (hereafter S16) to measure the SFR of the interacting system in the Q\,1436-0051 field (their object 6 and 7, see Section \ref{sec:NIRphot} for more information). They cannot resolve the flux from the interacting components. Using Salpeter IMF, they therefore report a combined SFR, which they split according to photometric flux-ratios, giving \mbox{$\mathrm{SFR}_{\mathrm{[\ion{O}{ii}]}}^{\mathrm{S16}}(\mathrm{obj}~6)=26~\mathrm{M}_\odot~\mathrm{yr}^{-1}$} and \mbox{$\mathrm{SFR}_{\mathrm{[\ion{O}{ii}]}}^{\mathrm{S16}}(\mathrm{obj}~7)=22~\mathrm{M}_\odot~\mathrm{yr}^{-1}$}. Our FORS2 spectrum is able to resolve the individual components of the interacting system (see Appendix \ref{appendix_Q1436-0051}), giving a $\mathrm{SFR}_{\mathrm{H}\beta} = 0.48 \pm 0.10~\mathrm{M}_\odot~\mathrm{yr}^{-1}$ of the confirmed host (see Table \ref{tab:results}). The strong contrast between the $\mathrm{SFR}_{\mathrm{[\textsc{O ii}]}}^{\mathrm{S16}}$ and our $\mathrm{SFR}_{\mathrm{H}\beta}$ diagnostic makes this field a unique case, granting additional analysis. Our FORS2 spectrum does not cover the [\textsc{O ii}] doublet. We therefore match the continuum light in the calibrated Magellan II spectrum used for the S16 analysis \citep[][and private communication with Lorrie Straka]{Straka2016} to the combined i-band magnitude reported in \cite{Rao2011}. Under the assumption that the S16 flux-level of $\times 10^{-17}~\mathrm{ergs}~\mathrm{s}^{-1}~\mathrm{cm}^{-2}$ is in fact $\times 10^{-18}~\mathrm{ergs}~\mathrm{s}^{-1}~\mathrm{cm}^{-2}$, the continuum fluxes are consistent. With this rescaling of the flux level, and converting to a Chabrier IMF, we recover a $\mathrm{SFR}_{\mathrm{[\ion{O}{ii}]}}^{\mathrm{This~work}} = 1.02 \pm 0.07~\mathrm{M}_\odot~\mathrm{yr}^{-1}$, which we also report in Table \ref{tab:results}. This value is consistent with our FORS2 $\mathrm{SFR}_{\mathrm{H}\beta}$ measurement.

%--------------------------------------------------------------------

\subsection{Metallicity}
    \label{sec:metallicity}
    We infer the metallicity of the absorbing galaxy in two independent ways based on (i) strong-line diagnostics with deredened line fluxes according to Section \ref{sec:EBV} where possible; and (ii) by applying a metallicity gradient correction to the known absorption metallicity. Despite being interchangeable, we distinguish between absorption and emission based measurements by referring to the metallicity ($[\mathrm{M}/\mathrm{H}]$) and Oxygen abundance $12+\log(\mathrm{O}/\mathrm{H})$, respectively.
    
    Emission-line fluxes in $[\mathrm{\ion{O}{ii}}]$, $\mathrm{H}\beta$ and $[\mathrm{\ion{O}{iii}]}$ enable us to calculate specific line-ratios which correlate with metallicity \citep{Pagel1979, Maiolino2008}. In the following, we report the results based on the $\mathrm{R}_{23}$, $\mathrm{O}_{32}$, $\mathrm{O}_2$ and $\mathrm{O}_3$ line-ratios, defined as
    \begin{equation}
        \label{eq:diagnostics}
        \begin{aligned}
            \mathrm{R}_{23} &= (f_{\mathrm{[\textsc{O ii}]} \lambda \lambda 3727,3729} + f_{\mathrm{[\textsc{O iii}]} \lambda \lambda 4959,5007}) / f_{\mathrm{H}\beta} \\
            \mathrm{O}_{32} &= f_{\mathrm{[\textsc{O iii}]} \lambda \lambda 4959,5007} / f_{\mathrm{[\textsc{O ii}]} \lambda \lambda 3727,3729} \\
            \mathrm{O}_2 &= f_{[\mathrm{\textsc{O ii}}]\lambda \lambda 3727,3729} / f_{\mathrm{H \beta}} \\
            \mathrm{O}_3 &= f_{[\mathrm{\textsc{O iii}}]\lambda 5007} / f_{\mathrm{H \beta}},
        \end{aligned}
    \end{equation}
respectively. These line-ratios are sensitive to the nebular conditions that govern the individual line-strengths, and observed line-ratios are often compatible with multiple abundances \citep{Osterbrock2006}. We therefore assume the standard $\mathrm{R}_{23}$ diagnostic, invoking the $\mathrm{O}_{32}$ ratio to account for ionisation corrections and break the degenerate solutions in abundance from a characteristic double-branched diagnostic. For systems with limited line-coverage, we report solutions based on the $\mathrm{O}_2$ and $\mathrm{O}_3$ line-ratios. Where these line-ratios give double-valued abundance-solutions, we report the abundance which closest matches the value inferred from a $z=0.7$ mass-metallicity relation \citep[][hereafter M08]{Maiolino2008}. For internal consistency, these diagnostic line-ratios are converted to oxygen abundances assuming the M08 calibration (see Table \ref{tab:results}). The reported uncertainties reflect the uncertainties in derived line-ratios, excluding the internal scatter present in the M08 relations. For notes on individual objects, see Appendix \ref{sec:notes}.

The use of different strong-line diagnostics, each displaying its own internal scatter, makes it difficult to analyse the sample in a homogenous way. By combining SED stellar masses with absorption-and emission line metallicities, \cite{Christensen2014} reinterpreted $C_{[\text{M/H}]}$ (see Section \ref{sec:intro}) as an average metallicity gradient $\Gamma$ acting over the impact parameter $b$ connecting the absorption- and emission measurements, $C_{[\text{M/H}]}=\Gamma b$. They retrieved a mean linear metallicity gradient $\langle \Gamma \rangle = - 0.022 \pm 0.004~\mathrm{dex~kpc}^{-1}$. In Section \ref{sec:metallicity_gradient} we re-examine the analysis including the results from this work, which gives a value $\langle \Gamma \rangle = - 0.022 \pm 0.001~\mathrm{dex~kpc}^{-1}$. Applying this updated correction factor and a 12 kpc truncation radius (see Section \ref{sec:metallicity_gradient}) to the absorption-metallicities allows us to infer standardised metallicity measurements in emission for each absorbing galaxy as
    \begin{equation}
        \label{eq:metallicity}
        [\mathrm{M}/\mathrm{H}]_{\mathrm{em}} = \left [\mathrm{M}/\mathrm{H}]_{\mathrm{abs}} + \langle \Gamma \rangle b~ \right \vert _{b \leq 12~\mathrm{kpc}} ~.
    \end{equation}
The calibration of Equation \ref{eq:metallicity} is tied to the M08 MZ-relation, which in turn uses the \cite{Kewley2002} functional form of the $\mathrm{R}_{23}$ and the $\mathrm{O}_{23}$ diagnostics to convert observed line-ratios to metallicity. We use zinc (Zn) as a tracer element of the absorption-metallicity where possible, since it is minimally depleted. Where only iron (Fe) or chromium (Cr) measurements exist, we include a standard depletion and/or $\alpha$-enhancement correction to the absorption metallicity, $\mathrm{[M/H]}_{\mathrm{abs}}=\mathrm{[Fe/H]}+0.3$\,dex \citep{Rafelski2012}. We note, however, that such a constant correction does not account for metallicity-dependent depletion, and \cite{DeCia2018} suggest that even Zn can be subject to marginal dust depletion, albeit to a lesser extent that for Fe. For this reason, the reported Zn-measurements will represent a lower limit to the metallicity measurements. 

%--------------------------------------------------------------------

\section{Discussion}
\label{sec:analysis}

In the following, our sample will be discussed in relation to other luminosity- and absorption selected samples alike. Luminosity-selected data is not illustrated for individual galaxies. We strictly adhere to plotting the scaling relations derived from such samples. The luminosity-selected MZ-relations are taken from M08, whereas the luminosity-selected star-forming main sequences are taken from \cite{Whitaker2014}. We refer to each relation's reference with the abbreviation M08 and W14, respectively. Likewise, we refer to \cite{Moller2013} as M13 - in particular when referring to the predicted M13 MZ-relation. Where relevant parameters are known, individual galaxies in the absorption-selected reference sample are plotted. The comparison sample refers to spectroscopically confirmed systems in \cite{Krogager2012} and \cite{Christensen2014}, referred to as K12 and C14, respectively. In Section \ref{sec:SFR} we extend the comparison sample to include the objects in \citet{kanekar2018}.

\subsection{Impact parameter as probe of average size}
\label{sec:impparam_distrib}
\begin{figure*}
    \centering
    \subfigure{\label{fig:b_distrib_left}\includegraphics[width=0.45\textwidth]{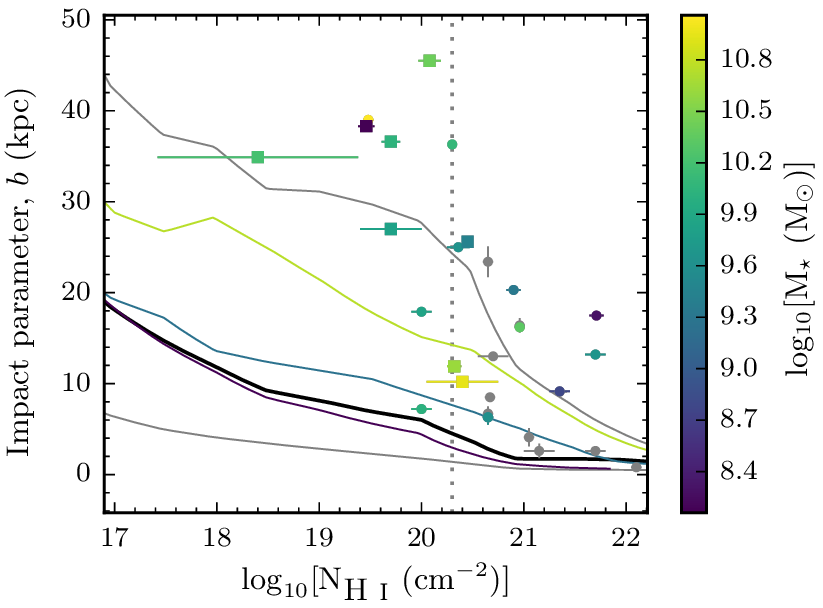}}
    \subfigure{\label{fig:b_distrib_right}\includegraphics[width=0.45\textwidth]{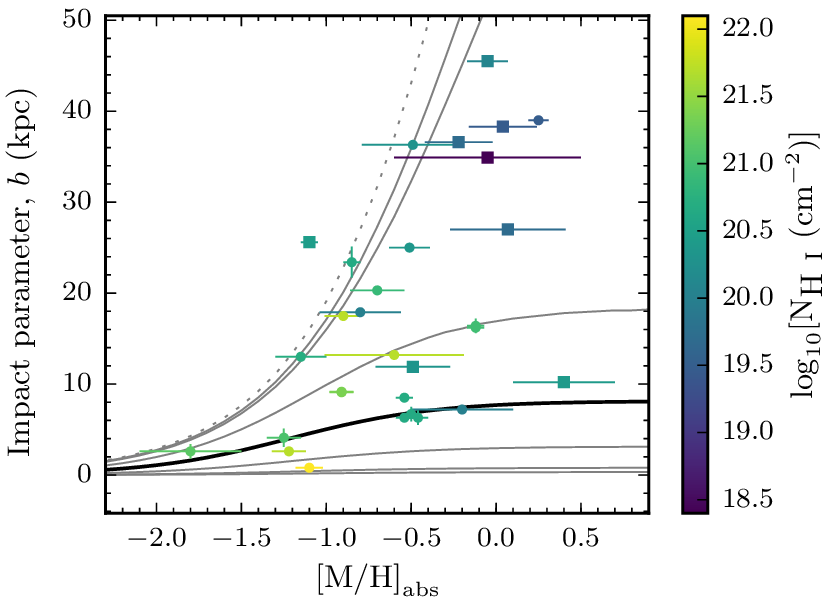}}
    \caption{\textit{left}: The distribution of impact parameter plotted as a function of the column density $\log _{10} [\mathrm{N}_{\mathrm{\ion{H}{i}}}~(\mathrm{cm}^{-2})]$, color-coded by stellar mass. Circles refer to data from literature. Our new identifications are plotted as squares. Grey data points represent spectroscopically confirmed absorbing galaxies without stellar mass estimates. The black line represents the median- and the grey lines represent the 1$\sigma$ scatter in simulated distributions \citep{Rahmati2014} at $z=2$. Color-coded lines represent the median impact parameters for those stellar mass bins, from the same simulations. \textit{Right}: The distribution of impact parameters as a function of the absorber metallicity $\mathrm{[M/H]_{\mathrm{abs}}}$, colour-coded by column density. The black line marks the median- and the grey lines mark the $1\sigma$, $2\sigma$ and $3\sigma$ contours based on the model-predictions from \cite{Krogager2017}. The grey dotted line marks the distribution envelope.}
    \label{fig:b_distrib}
\end{figure*}

The impact parameter, $b$, measures the projected distance between the luminosity-centre and a random quasar sightline piercing the gaseous component of a galaxy. Therefore, the distribution of impact parameters conveys information on the average spatial scales and sizes of the gaseous \ion{H}{i} regions probed, having folded in detection- and strategy-biases. In Figure \ref{fig:b_distrib}, we therefore plot the distribution of $b$ as a function of $\log _{10}[\mathrm{N}_\mathrm{\ion{H}{i}}~(\mathrm{cm}^{-2})]$ (\textit{left}) and as a function of $[\mathrm{M/H}]_\mathrm{abs}$ (\textit{right}). The data refers to damped absorbers with spectroscopically confirmed hosts in the reference sample (circles) with the added objects from this work (squares). Our sample effectively extends the observed distributions of Figure \ref{fig:b_distrib} towards (i) lower redshifts; (ii) lower \ion{H}{i} column-densities; and (iii) higher absorption-line metallicities.

Building on the results of previous studies (see Section \ref{sec:intro}), the left panel of Figure \ref{fig:b_distrib} conforms to the idea that sub-DLAs and DLAs trace different relations to their hosts; the former having higher scatter- and higher mean impact parameters than the latter, on average. Similar conclusions were reached by \cite{Rahmani2016} who found a transition region at $\sim 20$\,kpc, above which no $\log _{10}[\mathrm{N}_\mathrm{\ion{H}{i}}~(\mathrm{cm}^{-2})] > 21.0$ absorbing galaxy was detected. We find a similar value, which for the combined sample becomes $b \sim 25~\mathrm{kpc}$. \cite{Rao2011} find that DLAs are statistically closer to the candidate host sightline than sub-DLAs by a factor of two, with $\langle b \rangle_{\mathrm{R11, DLA}} = 17.4$\,kpc and $\langle b \rangle_{\mathrm{R11, sub-DLA}} = 33.3$\,kpc, respectively. This is remarkably consistent with our findings, which give median impact parameters of $\langle b \rangle_{\mathrm{DLA}} = 11.1$\,kpc and $\langle b \rangle_{\mathrm{sub-DLA}} = 35.8$\,kpc. These results suggest a characteristic scale-length associated with  the column-density distribution of gas in absorption-selected galaxies, and may be connected to the fall-off in gas density with radius.

We over-plot the median (black) and $1\sigma$ (grey) lines of the impact parameter distribution from the \cite{Rahmati2014} simulations. We colour-code the data to the SED $\log _{10}[\mathrm{M}_\star~(\mathrm{M}_\odot)]$ (the point is flagged grey if no stellar mass value exists) and over-plot the median curves for $\log _{10}[\mathrm{M}_\star~(\mathrm{M}_\odot)]$ bins of $7.0-8.5$; $8.5-10.0$; and $10.0-11.5$ \citep{Rahmati2014}, in representative colours. This allows us verify whether the observed distribution in the data is captured by the simulations, and to assess whether the retrieved column-density scale-length is mass-dependent. With the current small sample, we find no such dependence.

Overall, the observed scatter is larger, and the median impact parameter increases sharper with decreasing \ion{H}{i} column density, relative to the simulations. One suggestion is that this difference is driven by false-positive host galaxy identifications \citep{Rahmati2014}. However, such a false-positive identification rate would have to be sensitive to the column-density of the absorber -acting stronger on lower column density absorbers in order to reconcile the observations with simulations. That sub-DLAs are located at large impact parameters does not make them less likely to be the host, an issue discussed in \cite{Meiring2011}. In addition, the imposed cut in metallicity limits this bias if absorbing galaxies follow a mass-metallicity relation (M13).

Instead, the interpretation is complicated by \textit{sample pre-selection}; \textit{detection-bias}; \textit{strategy-bias}, and there may be a \textit{size-evolution with redshift}. Fundamentally, an absorption-selection introduces a bias towards gaseous systems with large neutral gas cross-sections. In combination with a lower cut on absorption metallicity, this pre-selects towards large, massive, bright galaxies independent of the absorber $\ion{H}{i}$ column density, biasing the observations towards large impact parameters. This selection is enhanced by the brightness of the background quasar, which prevents the detection of low luminosity systems -especially at low impact parameters if the region below the quasar trace is not scanned with an SPSF analysis (see Section \ref{sec:SPSF}). In addition, our $z<1$ data is a long-slit follow-up of systems with photometric candidate hosts, whereas the $z \geq 2$ galaxies are identified from $\mathrm{Ly}\alpha$ emission in the $\mathrm{Ly}\alpha$ trough or from strong optical emission lines found along any of the three slit position angles configured to cover as much of the region around the quasar as possible, without any pre-selection on candidate host \citep{Fynbo2010, Krogager2017}. Intuitively the latter observing strategy leads to a higher effective exposure time in the slit-coverage overlap, and therefore to a preferential identification of hosts at low impact parameters. 

This strategy-bias is quantified in the right panel of Figure \ref{fig:b_distrib}, which shows how the impact parameters (circles and squares) are distributed relative to the modelled expectation \citep[grey contours;][]{Krogager2017} as a function of absorption metallicity. The data is now colour-coded to the \ion{H}{i} column density, with an expected correlation assuming the existence of a luminosity-metallicity relation for damped absorbers (M13), and that luminosity correlates with size of the absorbing galaxy (K12). In their paper, \cite{Krogager2017} show a remarkable statistical agreement; the $1\sigma$ model contours encompassing $69\%$ of their data. This implies that the strategy bias is effectively negligible in their DLA sample, relative to the pre-selection towards large systems. 

It is remarkable that the model, developed for $z\sim 2-3$ DLAs, has an envelope that encompasses $96\%$ of the spectroscopically confirmed systems (i.e. $27/28$ systems, with one outlier) extending towards sub-DLAs and to $z\sim 0.7$. To investigate this statistically, we compute the fraction of systems locked into the $1\sigma$-, $2\sigma$-, $3\sigma$- model contours, for which we find $46\%$; $82\%$; and $89\%$ of the combined data, respectively. This factor two increase between the $1$- and $2\sigma$ contours indicates that that model, does not statistically match a combined sample of sub-DLAs and DLAs. This is supported by the \ion{H}{i} column-densities, which show that the factor of two increase is attributed to the more metal-enriched ($\mathrm{[M/H]}_{\mathrm{abs}} \geq -0.5$) sub-DLAs. 

%--------------------------------------------------------------------
  
\subsection{Drivers of the apparent $\mathrm{M}_\star - \mathrm{N}_{\mathrm{\textsc{Hi}}}$ correlation}

\begin{figure}
    \centering
    \includegraphics[width=0.5\textwidth]{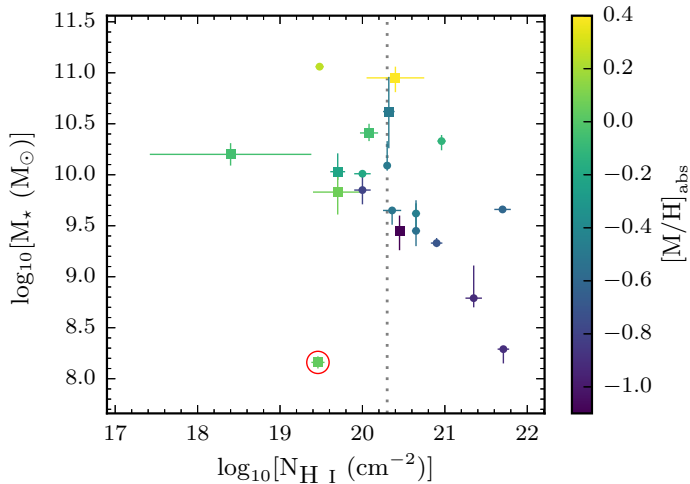}
    \caption{SED stellar mass of absorbing galaxies as a function of the absorber neutral hydrogen column density, colour-coded by the absorption metallicity. The dotted line marks the traditional distinction between sub-DLA and DLA at $\log _{10}[\mathrm{N}_{\mathrm{\textsc{H i}}} (\mathrm{cm}^{-2})] = 20.3$. The object encircled in red represents the outlier from Q\,1209$+$107 PA2.}
    \label{fig:Mstar_NHI}
\end{figure}

It has been suggested that sub-DLAs and DLAs arise in galaxies of different masses \citep{Kulkarni2010, Meiring2011} - a suggestion which may alleviate the statistical tension between sub-DLAs and DLAs in the $b$-$[\mathrm{M/H}]_\mathrm{abs}$ model expectation (see Section \ref{sec:impparam_distrib} and Figure \ref{fig:b_distrib}). To address this question, in Figure \ref{fig:Mstar_NHI} we plot the stellar mass of spectroscopically identified hosts as a function of the \ion{H}{i} column density of the associated quasar absorber, colour-coded to the absorption-metallicity, $\mathrm{[M/H]_{\mathrm{abs}}}$.

The data suggests an anti-correlation, with higher-mass systems observed to host progressively lower column-density absorbers; and a secondary correlation between the absorption-metallicity and stellar mass, reminiscent of a mass-metallicity relation. But the observed anti-correlation could be a manifestation of multiple effects acting simultaneously; (i) a selection-bias against dusty -and therefore massive, metal-rich systems; and (ii) and the cross-section selection biasing the sample number-count to large, massive galaxies with relatively larger projected areas of lower-density gas; the combination of the two making the sampling of the underlying distribution mimic a correlation. 

We know that visual attenuation, $A_V$, which is used to parametrise the dust content and the overall reddening of a source is related to the total column of metals. The metal column is itself the product of the neutral hydrogen gas column density and the metallicity. Recent studies have shown that the metal column correlates with the visual attenuation, such that $\log_{10} (A_V/\mathrm{mag}) \propto \log_{10}\mathrm{N}_\mathrm{\textsc{H i}} + [\mathrm{M/H}]$ \citep[see][]{Zafar2013,Zafar2018}. Such a preferential dust-bias against high column-density absorbers is therefore consistent with the observed column-density dependent cutoff in stellar mass that we observe in the data, and can -at least in part, be explained by a dust bias against red quasars.

Recent work from the High $A_V$ Quasar survey \citep[HAQ; ][]{Fynbo2013b, Krogager2015,Zafar2015} - and the extended HAQ \citep[eHAQ; ][]{Krogager2016, Fynbo2017} survey suggest that a traditional quasar selection is indeed biased against reddened quasars. This is consistent with \cite{Noterdaeme2015} who found that the high column-density ($\log_{10} \mathrm{N}_\mathrm{\textsc{H i}} \sim 22$), high metallicity ($\gtrsim 1/10$ solar) DLAs induce a colour-change in the background quasar caused by both dust-  and hydrogen absorption. This propagates into a dust-bias which preferentially acts on high $\mathrm{N}_{\mathrm{\textsc{H i}}}$, massive, metal-rich, dusty galaxies selected against quasar sight-lines.

%--------------------------------------------------------------------

\subsection{The mass-metallicity relation}
\label{sec:mass-metallicity}

\begin{figure}
    \centering
    \subfigure{\label{fig:MZa}\includegraphics[width=0.5\textwidth]{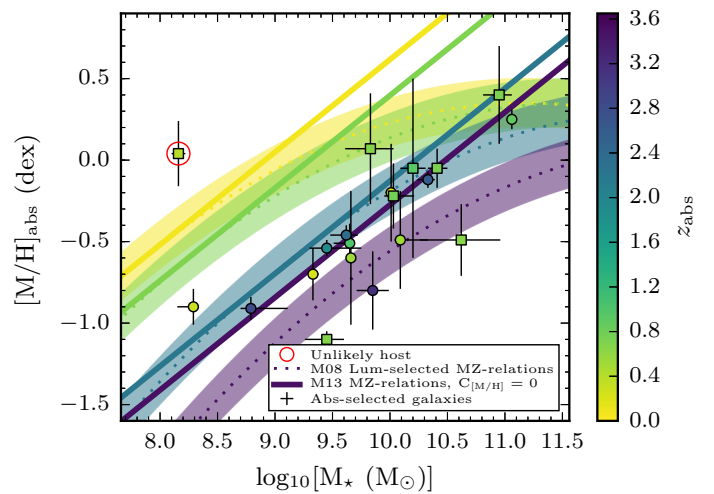}}
    \subfigure{\label{fig:MZb}\includegraphics[width=0.5\textwidth]{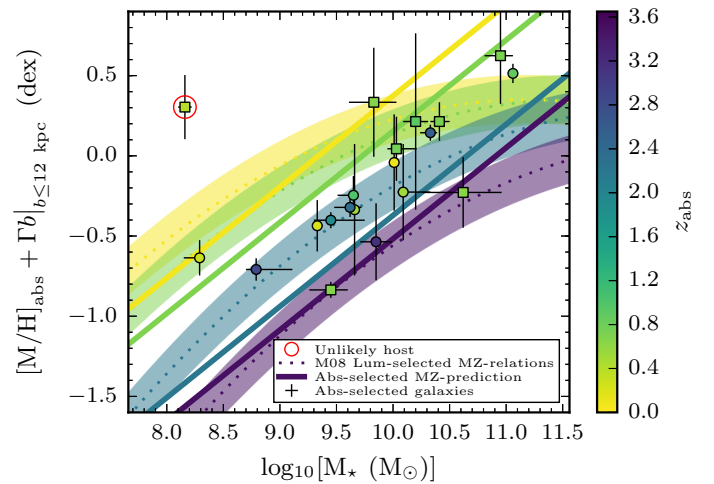}}
    \caption{The metallicity as a function of stellar mass for absorption-selected galaxies, colour-coded by redshift. Squares indicate the results of this work, whereas circles refer to the absorption-selected reference sample. The M08 mass-metallicity relations based on luminosity-selected samples, and the M13 predicted mass-metallicity relations for absorption-selected samples have been overplotted at four representative redshifts. The \textit{top} panel shows the results assuming the metallicity measured in absorption. The \textit{bottom} panel shows the results with an impact-parameter dependent correction factor and a gradient cut-off at at truncation radius of 12\,kpc. The object encircled in red represents the outlier from Q\,1209$+$107 PA2.}
    \label{fig:mass-metallicity}
\end{figure}

We investigate how the absorption-selected galaxy population is distributed relative to known mass-metallicity (MZ) relations of luminosity-selected galaxies.  In Figure \ref{fig:mass-metallicity}, we plot two metallicity measurements as a function of the stellar mass. The data represents spectroscopically confirmed absorbing galaxies, with squares denoting the measurements in this work, and circles representing the compiled reference sample. We colour-code the data according to the redshift of each system, and overplot the M08 MZ relations at redshifts $z=0.07$, $z=0.7$, $z=2.2$, and $z=3.5$, matched in colour. Here, we also add a $0.15~\mathrm{dex}$ conservative estimate of the intrinsic scatter in each relation as the shaded regions \citep[][M08]{Kewley2008}.

In Figure \ref{fig:mass-metallicity} \textit{top} panel, we use the absorption metallicity as an indicator of the integrated emission metallicity. In addition, we convert the M13 functional form for $\mathrm{M}_\star$ into a mass-metallicity relation (M13 MZ-relation) for absorption-selected galaxies, setting $C_{\mathrm{[M/H]}}=0$. These are plotted for the same four redshifts as the M08 relations to give a fair comparison. Two observations can be made: (I) With the exception of $z=3.5$, the M08- and M13 MZ relations tangentially match in the near linear low-mass regime, but diverge towards larger masses. (II) The data fall systematically below both the M08 and the M13 MZ-relations at their respective redshifts. To quantify how well the models describe the data, we calculate the reduced chi-square statistic, $\chi^2_\nu$. For M08, we first interpolate the functional form to the redshifts of the data. Including the 0.15 dex intrinsic scatter in the relation, we find a value \mbox{$\chi^2_{\nu,~\mathrm{M08}} = 11.70$}. This can be compared to the corresponding M13 MZ-relation which, by including their intrinsic scatter of 0.38 dex yields a value \mbox{$\chi^2_{\nu,~\mathrm{M13}} = 3.21$}. We here note that whilst the M13 model is statistically a better fit to the data, this result is driven by a large internal scatter, and does not mitigate the systematically lower metallicities for a given stellar mass and redshift.

Following the prescription outlined in Section \ref{sec:metallicity}, in the \textit{bottom} panel of Figure \ref{fig:mass-metallicity}, we correct the absorption metallicity for the average metallicity gradient solved in Section \ref{sec:metallicity_gradient}. Local galaxies are found to have effective radii $\sim 6$\,kpc, with oxygen abundance gradients extending to around $\sim 2$ disc effective radii followed by flat gradients \citep[the CALIFA survey,][]{Sanchez2014}. Motivated by these results, in combination with the metallicity gradient we also apply a truncation-radius at $12~\mathrm{kpc}$ to the data. This effectively converts the absorption measurement into an indirect emission measurement of the metallicity at the galaxy luminosity centre.

It is worth noting that the M08 MZ-relations are calibrated to metallicity measurements averaged over the emission line regions of star forming galaxies. We can therefore ask whether applying a correction factor to the absorption metallicity will overestimate the luminosity-selected measurement, rendering the comparison meaningless? The individual surveys used to calibrate the M08 MZ relations at different redshifts have apertures of $0.75-1.3~\mathrm{arcsec}$ \citep[][$z\sim 0.7$]{Savaglio2005}; $0.76~\mathrm{arcsec}$ \citep[][$z\sim 2.2$]{Erb2006} and $0.75~\mathrm{arcsec}$ \citep[][$z\sim 3.5$]{Maiolino2008}. A source centred on -and filling the aperture will therefore have physical radii of the order $2.8 - 4.7~\mathrm{kpc}$; $3.6~\mathrm{kpc}$ and $3.0~\mathrm{kpc}$ for those representative redshifts, respectively. This gives a conservative estimate of the overshooting in the range \mbox{$r_\mathrm{aperture} \times \langle \Gamma \rangle \sim 0.06 - 0.10$ dex}. Recognising that this range is significantly below individual metallicity uncertainties, and that the exact overshooting is sensitive to the light-distribution across the aperture, we conclude that its net systematic effect on our conclusions are negligible.

In the lower panel, we also overplot the same M08 relations, and include $C_{\mathrm{[M/H]}}$ to the M13 predicted MZ-relation. To be consistent with the original functional form based on a pure DLA-sample, we reformulate the constant as $C_{\mathrm{[M/H]}}= \langle \Gamma \rangle \times \langle b \rangle _{\mathrm{DLA}}$, for $\langle \Gamma \rangle = 0.022~\mathrm{dex}~\mathrm{kpc}^{-1}$ (see Section \ref{sec:metallicity_gradient}), and $\langle b \rangle _{\mathrm{DLA}} = 11.1~\mathrm{kpc}$ (see Section \ref{sec:impparam_distrib}). Having applied the corrections to the data and to the relations, we now observe that: (I) Rather than overlapping with the M08 relations at low stellar masses, including $C_\mathrm{[M/H]}$ renormalises the M13 MZ-relation and generates convergence at larger stellar masses, on average. (II) The systematic bias in the data towards lower metallicity for a given stellar mass and redshift bin has been reduced. These conclusions are supported by recalculating the reduced chi-square statistics, which give values of \mbox{$\chi^2_{\nu,~\mathrm{M08}} = 7.95$} and \mbox{$\chi^2_{\nu,~\mathrm{M13}} = 0.98$} by replacing the intrinsic scatter from M13 with the updated value of 0.32 (see Section \ref{sec:metallicity_gradient}). 

We therefore conclude that the absorber based MZ-relation provides a better fit to the data. We recognise that this fit is dominated by a large intrinsic scatter that prevents us from discerning any clear redshift evolution. Despite this, the inclusion of a mean metallicity gradient is physically motivated, and has a significant effect on the $\chi^2_\nu$-statistic, reducing it by a factor $\sim 3$. This suggests that we can statistically predict global properties from local measurements.

%--------------------------------------------------------------------

\subsection{Absorbing galaxies probing sub-main-sequence star-formation}
\label{sec:SFR}

\begin{figure}
    \centering
    \includegraphics[width=0.5\textwidth]{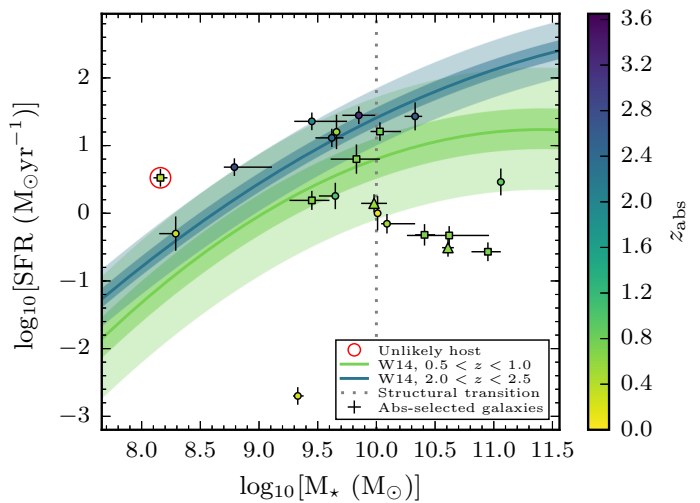}
    \caption{\label{fig:SFR} The star-formation rate as a function of stellar mass for spectroscopically confirmed absorbing galaxies, colour-coded by redshift. Squares indicate the results of this work; circles refer to work by others. Triangles represent CO detected galaxies. For reference, we plot the \cite{Whitaker2014} redshift-dependent star-forming main sequence for two representative redshift ranges together with their $1\sigma$- and $3\sigma$ scatter, represented by the dark- and light- shaded regions centred on each relation, respectively. See text in Sec. \ref{sec:SFR} for more details. The object encircled in red represents the outlier from Q\,1209$+$107 PA2.}
\end{figure}

\cite{Moller2018} and \citet{kanekar2018}, based on a recent ALMA survey of molecular gas \citep{Neeleman2016, Moller2018}, try to characterise the relation between the neutral- and molecular gas content and the SFR in absorption-selected galaxies. Based on a sample of two systems, the former suggests that absorption-selected systems sample a missing phase in galaxy evolution; a `post-starburst' phase, characterised by low star-formation and large gas- and molecular fractions \citep{Moller2018}. This is driven by the preceding onset of a starburst phase that drives molecular- and neutral gas to large distances, making such systems susceptible to the \ion{H}{i} cross-section selection. In the latter, \cite{kanekar2018} from a sample of seven, argue that absorption-selected galaxies are consistent with the star-forming main-sequence; but that the systems retain large molecular masses at all redshifts. The insensitivity of the SFR on molecular gas mass is interpreted as evidence for low molecular gas-densities which inhibit the molecular-gas from converting into stars.

Here, we do not consider CO-detections. This allows us to look at a larger data-set to quantify the absorbing galaxy population relative to the star-formation main sequence as a whole. In figure \ref{fig:SFR}, we show how the sample of absorbing galaxies fall relative to the main sequence of star-formation. Square symbols represent the results of this work (SED-based $\mathrm{M}_\star$ and extinction-corrected SFRs, as reported in Table \ref{tab:results}); circles refer to literature samples. For completeness, we add to our sample the two systems (B\,1629+120g and J\,0058+0155g) from \citet{kanekar2018}, not covered by the C14 sample or this study. These two systems are represented by triangles. For galaxies with multiple SFR-tracers ([\ion{O}{ii}] and $\mathrm{H}\beta$), we assign preference to the $\mathrm{H}\beta$ measurement, since this is a direct measure of the SFR based on recombination theory. We note however that in general, the [\ion{O}{ii}] SFRs show good agreement with the $\mathrm{H}\beta$ based measurements. In addition to the measurement errors reported in Table \ref{tab:results}, we include a $30\%$ uncertainty to each SFR to account for scatter between the $\mathrm{H}\beta$ and the [\ion{O}{ii}] diagnostic calibrations \citep{Kennicutt1998, Kewley2004}. We then colour-code the absorbing galaxies by redshift. For reference, we over-plot the star-forming main-sequences at redshift ranges $0.5 < z < 1.0$ and $2.0 < z < 2.5$ for a Chabrier IMF \citep{Whitaker2014}. We have matched the colours for each relation to representative redshifts of $z=0.75$ and $z=2.25$, respectively. In addition, we include the $1\sigma$- and $3\sigma$ dispersions as the dark- and light shaded regions around each of the correlations. These are estimated based on a reported scatter of 0.14 dex for $z\sim 2$ and 0.30 dex for $z<1$ \citep{Whitaker2015}.

The high-redshift ($z \sim 2$) absorbing galaxies are consistent within the $3\sigma$ scatter of the star-forming main sequence at that redshift, whereas a significant fraction of systems at low ($z\lesssim 1$) redshifts suggest sub-main-sequence star-formation. Despite individual systems showing large scatter, we note that this bias appears significant and real, since the data are not stochastically distributed around the main sequence. With the current sample, we therefore ask whether this suppression, if real, is \textit{evolutionary} or \textit{inherent} to the nature of absorption-selected galaxies?

Despite the temptation to see the results as indicative of an evolutionary transition, with galaxies ``breaking off'' the star-forming main sequence to become suppressed/quenched at some characteristic redshift $z \sim 0.7$ in line with the CO detected DLA galaxies in \citet{kanekar2018}, this is not supported by our data. A majority of the systems with suppressed SFRs lie at the high-mass end of the distribution, whereas the majority of systems at lower stellar masses are consistent with the $3\sigma$ scatter. Empirically, this suggests that suppressed SFRs are inherent to absorption-selected galaxies above stellar mass \mbox{$\log _{10}\mathrm{M}_\star ~(\mathrm{M}_\odot) \gtrsim 10$}, indicated by the grey dotted line in Figure \ref{fig:SFR}. 88\% (7/8 systems) with $\log _{10}\mathrm{M}_\star ~(\mathrm{M}_\odot) \gtrsim 10$ fall below their respective redshift-relation. Out of these, 63\% (5/8 systems) are inconsistent with the $3\sigma$ scatter. Despite the low number of systems at high-z, this interpretation is consistent with the Q\,0918+1636 system ($z_\mathrm{abs} = 2.583$, $\log _{10}\mathrm{M}_\star ~(\mathrm{M}_\odot) \sim 10.3$), with a single outlier in Q\,0738+313 ($z_\mathrm{abs} = 0.221$, $\log _{10}\mathrm{M}_\star ~(\mathrm{M}_\odot) \sim 9.3$) displaying suppressed SFR at lower mass.

The natural question to ask is whether this suppression can be explained by selection effects or a dust bias. Indeed, pre-selecting against HAQ galaxies \citep{Fynbo2013b, Krogager2015, Zafar2015}, large fractions of obscured star formation and lack of applying extinction corrections to the SFR will systematically deplete the high mass, high star formation domain of Figure \ref{fig:SFR}. \cite{Ma2015} show that massive ($10.3 \leq \log _{10}[\mathrm{M}_\star ~(\mathrm{M}_\odot)] \leq 11.5$) , dusty galaxies are usually highly star forming, with values of $510 - 4800 ~\mathrm{M}_\odot ~\mathrm{yr}^{-1}$. The extinction corrected SFRs for our galaxies (see Table \ref{tab:results}) are low, which suggests that the detected galaxies on average are not very dusty. Based on the high fraction of identified hosts, a dust bias therefore does not appear to affect this study.

Albeit difficult to motivate why this mass-scale would be important in the context of galaxy evolution, we therefore suggest that the current data supports the claim that suppressed SFR in absorption-selected galaxies is mass-dependent, at a mass-scale of $\log _{10}[\mathrm{M}_\star ~(\mathrm{M}_\odot)] \sim 10$, and not evolutionary.

%--------------------------------------------------------------------

\subsection{The existence of an average metallicity gradient}
\label{sec:metallicity_gradient}

Cosmological hydrodynamic simulations such as the Galaxies-Intergalactic Medium Calculation \citep[GIMIC,][]{Crain2009} predict systematic flattening of metallicity-gradients in star-forming disc-galaxies with redshift, from $z=2$ to $z=0$. In such simulations, this flattening is caused by declining inflows of pristine gas; enrichment by \textit{in situ} star-formation; and the redistribution of metal-enriched gas from the inner to outer disc, leading to progressive outer disc enrichment. Observations of spatially resolved star-forming galaxies are comparable to these results \citep{Swinbank2012}. Likewise, \cite{Stott2014} suggest that the sSFR, parametrising how intensely a galaxy is forming stars, drives the observed differences in metallicity gradients. In their framework, observations of high-sSFR systems associated with metal-poor centres and flat/inverted metallicity-gradients are a consequence of mergers \citep{Rupke2010} and cold flows \citep{Dekel2009, Cresci2010}. Similar results have also been reported by \cite{Queyrel2012}, who finds examples of positive gradients in $z=1.2$ star-forming galaxies. These mechanisms will channel more pristine gas into the galaxy central regions, providing fuel and triggering intense star-formation that increases the sSFR whilst diluting the metallicity in the galactic cores. This leads us to expect a synergy between the metallicity-gradient's observed dependence on sSFR and redshift, consistent with the fundamental metallicity relation for which galaxies at fixed stellar mass are progressively metal poor for increased SFR \citep{Mannucci2010}. 

To see whether we can resolve such evolutionary diagnostics, we first determine individual metallicity-gradients $\Gamma_i~[\mathrm{dex~kpc}^{-1}]$. We recognise, however, that there are important differences between gradients measured in emission on spatially resolved systems to those measured by combining emission with absorption measurements. The derived oxygen abundances reported in Table \ref{tab:results} as well as those from literature use different emission-line diagnostics, and therefore represent a heterogenous sample. However, they are re-calibrated to M08 diagnostics, and should therefore be comparable. Additionally, absorption metallicities sample pencil-beams and therefore trace the local metallicity, typically offset from the galaxy centre. In slit spectra however, a spectroscopic emission-line metallicity represents the mean metallicity across the slit width. For a galaxy at $z=0.7$ (a representative redshift of our sample), a slit-width of $\theta = 1.31~\mathrm{arcsec}$ will yield a metallicity-estimate averaged over spatial scales of $\sim 9.5~\mathrm{kpc}$. We therefore caution against a direct combination of absorption and emission metallicities, since they likely probe different enrichment histories and sample local and global enrichment time-scales, respectively. 

To determine $\Gamma_i$ homogeneously, we instead use the relation $\Gamma_i = C_{\mathrm{[M/H]},i} / b_i$ (C14). $C_{\mathrm{[M/H]},i}$ refers to the predicted metallicity parameter (M13, their eq. 6), evaluated for a system $i$ assuming its SED stellar mass:
\begin{equation}
    C_{\mathrm{[M/H]},i} = \frac{\log _{10} [\mathrm{M}_{\star,i}^{\mathrm{SED}} (\mathrm{M}_\odot)]}{1.76} - [\mathrm{M}/\mathrm{H}]_{\mathrm{abs},i} - 0.35z_{\mathrm{abs},i} - 5.04~,
    \label{eq:CHM}
\end{equation}
and $b_i$ refers to the measured impact parameter. We then determine an empirical expectation value of the average metallicity gradient, $\langle \Gamma \rangle$, adopting the $C^2_{\mathrm{dof}}$-minimisation method used in M13 and C14. Recognising that Equation \ref{eq:CHM} will have an intrinsic/native scatter, $\sigma _{\mathrm{nat}}$, we desire the combination [$\sigma_{\mathrm{nat}}$, $\langle \Gamma \rangle$] which minimises $\sigma_{\mathrm{nat}}$ and for which the $C^2_{\mathrm{dof}}$-statistic is unity. In effect, we seek the solution to
\begin{equation}
    0 = \Bigg( \frac{1}{\mathrm{dof}} \sum_{i=1}^{N} \frac{\big(C_{\mathrm{[M/H]},i} - \langle \Gamma \rangle b_i\big)^2}{\sigma ^2_{\mathrm{nat}} + \sigma^2_{C_{\mathrm{[M/H]},i}}} \Bigg) - C^2_{\mathrm{dof}}~,
    \label{eq:MHgradient}
\end{equation}
where the degrees of freedom (dof) is $N-1$ and not $N-2$, since by construction we fix $\sigma_{\mathrm{nat}}$ to its minimum value. $\sigma_{C_{\mathrm{[M/H]},i}}$ is the propagated uncertainty from the absorption metallicity and from the stellar mass measurements. 

For our total FORS2 sample, we find a solution $[0.66\pm 0.05, 0.015\pm 0.001]$. Combining the new sample with C14, we find $[0.44\pm 0.01, 0.018\pm 0.001]$. These values are shallower than the findings of C14 who finds $\langle \Gamma \rangle _{\mathrm{C}14} = -0.022 \pm 0.004~\mathrm{dex~kpc^{-1}}$, and only marginally consistent to within the reported errors in our respective studies. We note however, that we reformulated a method that searched a pre-defined parameter-grid for the minimum $\sigma _{\mathrm{nat}}$ into a method seeking the roots of equation \ref{eq:MHgradient}. Having confirmed that the two methods yield consistent values on the same data, we exclude the probable misidentification in Q\,1209+107 PA2 (see Section \ref{sec:missidentification}). With this exclusion we retrieve an intrinsic dispersion $\sigma _\mathrm{nat} = 0.32\pm 0.01$ and an average gradient of $\langle \Gamma \rangle _{\mathrm{new}} = -0.022 \pm 0.001~[\mathrm{dex~kpc^{-1}}]$. This value is in perfect agreement with $\langle \Gamma \rangle _{\mathrm{C}14}$, and lends further support to the galaxy in Q\,1209+107 being an unlikely host.

In the same work, C14 solve for the direct gradients measured from the direct comparison of metallicity measurements in emission- and absorption line measurements, $([\mathrm{M/H}]_{\mathrm{em}} - [\mathrm{M/H}]_{\mathrm{abs}})/b_i$, for which they find a mean value of $-0.023 \pm 0.015~\mathrm{dex~kpc^{-1}}$. Similar work comparing direct measurements yield a median value of $-0.022~ \mathrm{dex~kpc^{-1}}$ \citep{Peroux2016a}, and performing a weighted mean on the three direct measurements reported in \citep{Rahmani2016} gives a metallicity gradient of $-0.04 \pm 0.01 \mathrm{dex~kpc^{-1}}$. We note however, that \cite{Rahmani2016} report a best-fit gradient to a compilation of all known measurements, excluding limits and non-detections of $-0.002 \pm 0.007~\mathrm{dex~kpc^{-1}}$. This would indicate an insignificant metallicity gradient - in contrast with our results, and may reflect the combination of heterogenous diagnostics and local- and global scale metallicity-measurements. \cite{Stott2014} retrieve an average gradient of $\Delta Z / dr = -0.002 \pm 0.007~\mathrm{dex~kpc}^{-1}$ based on 20 disc galaxies at $z=1$. \cite{Swinbank2012} recover metallicity gradients in discs ($\lesssim 10~\mathrm{kpc}$) that are negative or flat for seven spatially resolved galaxies at $z=0.84-2.23$, with an average gradient of $\Delta \log_{10} (\mathrm{O/H}) / \Delta R = -0.027 \pm 0.005~\mathrm{dex~kpc}^{-1}$.

Having robustly identified the existence of an average metallicity gradient in absorption-selected galaxies with a grid-based and a solver-based model alike, we now consider whether $\langle \Gamma \rangle$ correlates with galaxy evolution parameters. In particular, we desire to see whether $\langle \Gamma \rangle$ is sensitive to the inclusion of sub-DLAs; whether it shows a redshift-evolution; and whether it correlates with galaxy evolution parameters such as $\mathrm{M}_\star$ or sSFR. Such correlations have been found in luminosity-selected star-forming disc-galaxies at $z=1$ \citep{Stott2014}. Since the scatter in the sample is larger than the individual measurement errors, we proceed with the methodology developed, but replace $\langle \Gamma \rangle$ by a linear function of the form $\langle \Gamma \rangle = \theta _1 x + \theta _2$ in equation \ref{eq:MHgradient}. Here, $\theta$ represents the new parameters we seek to optimise, and $x$ is the independent variable.

\begin{table}[t!]
\caption{Retrieved metallicity-gradient correlations.}             
\label{tab:MHgrad_correlations}      
\centering
\small        
\begin{tabular}{l r}
\hline\hline
Independent Variable & $\Delta \Gamma / \Delta x$ \\
$(x)$ & $(\theta_1)$ \\
\hline
$\log _{10} [\mathrm{N}_\mathrm{\ion{H}{i}}~(\mathrm{cm}^{-2})]$ & $-0.016 \pm 0.003$ \\
$z_\mathrm{abs}$ & $0.07\pm 0.03$\\
$\log _{10} [\mathrm{M}_\star ~(\mathrm{M}_\odot)]$ & $-0.02 \pm 0.01$ \\
$\mathrm{SFR}~(\mathrm{M}_\odot~\mathrm{yr}^{-1})$ & $-0.003 \pm 0.001$ \\
$\log _{10}[\mathrm{sSFR}~ (\mathrm{yr}^{-1})]$ & $0.002 \pm 0.02$ \\
\hline
\end{tabular}
\end{table}

The resulting correlations are summarised in Table \ref{tab:MHgrad_correlations}. We find a $\sim 5\sigma$ correlation with the \ion{H}{i} column density, and a formal $3\sigma$ correlation with SFR. A correlation with \ion{H}{i} column density may be related to the ionisation-corrections applied to the absorption metallicity, which itself correlates inversely with \ion{H}{i} column density. Whether such ionisation corrections explain the observed correlation is beyond the scope of this paper.

If real, the correlation with SFR may suggest that absorption-selected galaxies have sSFRs which are driven by a different mechanism relative to luminosity-selected systems. Such systems show a correlation with sSFR, with slopes of $0.023 \pm 0.004$, driven by a correlation with $\mathrm{M}_\star$ of the order $-0.022 \pm 0.009$ \citep{Stott2014}. We note however, that with errors of 0.01 and 0.02 on the correlations with $\mathrm{M}_\star$ and sSFR, respectively, the data are consistent with the disc-galaxy correlations, but lack the number statistics to test the \cite{Stott2014} result.

%--------------------------------------------------------------------

\section{Conclusions}
\label{sec:conclusion}

In this paper we have spectroscopically confirmed seven galaxies harbouring damped absorption systems at redshift $z<1$. We additionally detect a low-mass, star-bursting galaxy at low impact-parameter in the Q1209+107 quasar field. Further analysis suggests that this is an unlikely host. This has two implications; (i) the galaxy hosts for the two damped absorbers in Q\,1209+107 are still unidentified; and (ii) the proximity to the line of sight suggests that the low-redshift ($z_\mathrm{abs} = 0.3930$) absorber is hosted by a galaxy that itself, belongs to a larger galaxy-group. We also report one non-detection (in the Q\,1209+107\,PA1 field), and a galaxy at the wrong redshift to be related to the absorbing system (in the Q\,2353-0028 field). With a conservative estimate, our FORS2 campaign has therefore had a success-rate of 70\%.

Combining spectroscopic data in emission with the known properties in absorption and deep multi-band photometry in the quasar fields, we have conducted an extensive analysis to characterise the absorbing galaxies. We summarise the results as follows:
\begin{itemize}
    \item Despite statistical differences, the observed distribution of impact parameters with \ion{H}{i} column-density and with absorption metallicity can be extended from DLAs to sub-DLAs; from high ($z\sim 2-3$) redshift to low ($z\sim 0.7$) redshifts; and to higher absorption metallicities. Our observations can be used to constrain the models further.
    \item We find no solid evidence for the claim that sub-DLAs and DLAs arise in galaxies of different masses. We argue the apparent correlation is driven by selection effects, including a dust-bias against massive metal-rich systems; and the fact that projected cross-section of low-column-density sub-DLA gas will be geometrically larger - thus favouring the detection of sub-DLAs in large galaxies.
    \item We find a significant mean metallicity gradient of $\langle \Gamma \rangle = -0.022 \pm 0.001~\mathrm{dex~kpc^{-1}}$ based on the absorber properties and an SED stellar mass. This value is consistent with literature values, but has a significantly reduced uncertainty achieved by avoiding the combination of different strong-line diagnostics (in emission) and by not directly combining measurements on local (absorption) and global (emission) spatial scales.
    \item Correcting the absorption-metallicities for a mean metallicity-gradient and assuming a physically motivated truncation-radius of $12$\,kpc, absorption-selected galaxies fall on top of predicted MZ-relations, with a significantly reduced scatter.
    \item Based on the current sample, absorption-selected galaxies with stellar masses $\log _{10}\mathrm{M}_\star ~(\mathrm{M}_\odot) \gtrsim 10$ show sub-main-sequence star-formation. Whereas this effect has been interpreted as an effect of probing galaxy-evolution phase and/or a transition from the nature of absorbing galaxies between high- and low redshifts, the fact that such a suppression is not seen at lower stellar masses, together with a low number of systems at high redshift, indicates a mass-dependence.
\end{itemize}

%--------------------------------------------------------------------

\begin{acknowledgements}
      This work was supported by grant ID \emph{DFF-4090-00079}. We would also like to extend our gratitude to Sandhya Rao and Lorrie Straka for sharing data; and to Jens-Kristian Krogager and Alireza Rahmati for sharing their model- and simulation predictions.
      
      This research made use of Astropy, a community-developed core Python package for Astronomy.
      
      In addition, it has made use of SDSS DR10. Funding for SDSS-III has been provided by the Alfred P. Sloan Foundation, the Participating Institutions, the National Science Foundation, and the U.S. Department of Energy Office of Science. The SDSS-III web site is http://www.sdss3.org/.

SDSS-III is managed by the Astrophysical Research Consortium for the Participating Institutions of the SDSS-III Collaboration including the University of Arizona, the Brazilian Participation Group, Brookhaven National Laboratory, Carnegie Mellon University, University of Florida, the French Participation Group, the German Participation Group, Harvard University, the Instituto de Astrofisica de Canarias, the Michigan State/Notre Dame/JINA Participation Group, Johns Hopkins University, Lawrence Berkeley National Laboratory, Max Planck Institute for Astrophysics, Max Planck Institute for Extraterrestrial Physics, New Mexico State University, New York University, Ohio State University, Pennsylvania State University, University of Portsmouth, Princeton University, the Spanish Participation Group, University of Tokyo, University of Utah, Vanderbilt University, University of Virginia, University of Washington, and Yale University.
\end{acknowledgements}

%--------------------------------------------------------------------

\bibliographystyle{aa.bst}
\bibliography{bibliography.bib}

%--------------------------------------------------------------------

\begin{appendix}

\section{Group analysis in Q1436-0051}
\label{appendix_Q1436-0051}

\begin{table*}[t!]
\caption{Galaxies at relatively small projected distances from the quasar Q1436-0051. The identified emission lines (see Figure \ref{fig:Q1436-0051_groupassociations}) suggests the presence of galaxy groups at the redshift of the absorbing systems redshifts (see Section \ref{sec:galassoc} for further discussion).}             
\label{tab:appendix_Q1436-0051}
\centering
\small        
\begin{tabular}{r l c}
\hline\hline
\multicolumn{3}{c}{Q1436-0051 Field Galaxy Identifiers}\\
$\theta$\tablefootmark{a} & $b$\tablefootmark{b} & $z_{\text{em}}^{\text{SPEC}}$\\
\hline
(-)5.0 & 36.7    & 0.7375 \\
(-)7.5 & 46.3    & 0.7382 \\
4.4    & 34.9    & 0.9289 \\
9.1    & 77.0    & 1.2632 \\
14.5   & 106.5   & 0.7379 \\
29.0   & 230.1   & 0.9273 \\
\hline 
\end{tabular}

\tablefoottext{a}{Un-binned angular distance relative to the quasar.}
\tablefoottext{b}{Projected distance relative to the quasar, measured at the spectroscopic redshift.}
\end{table*}

\begin{figure*}
    \centering
    \includegraphics[width=\textwidth]{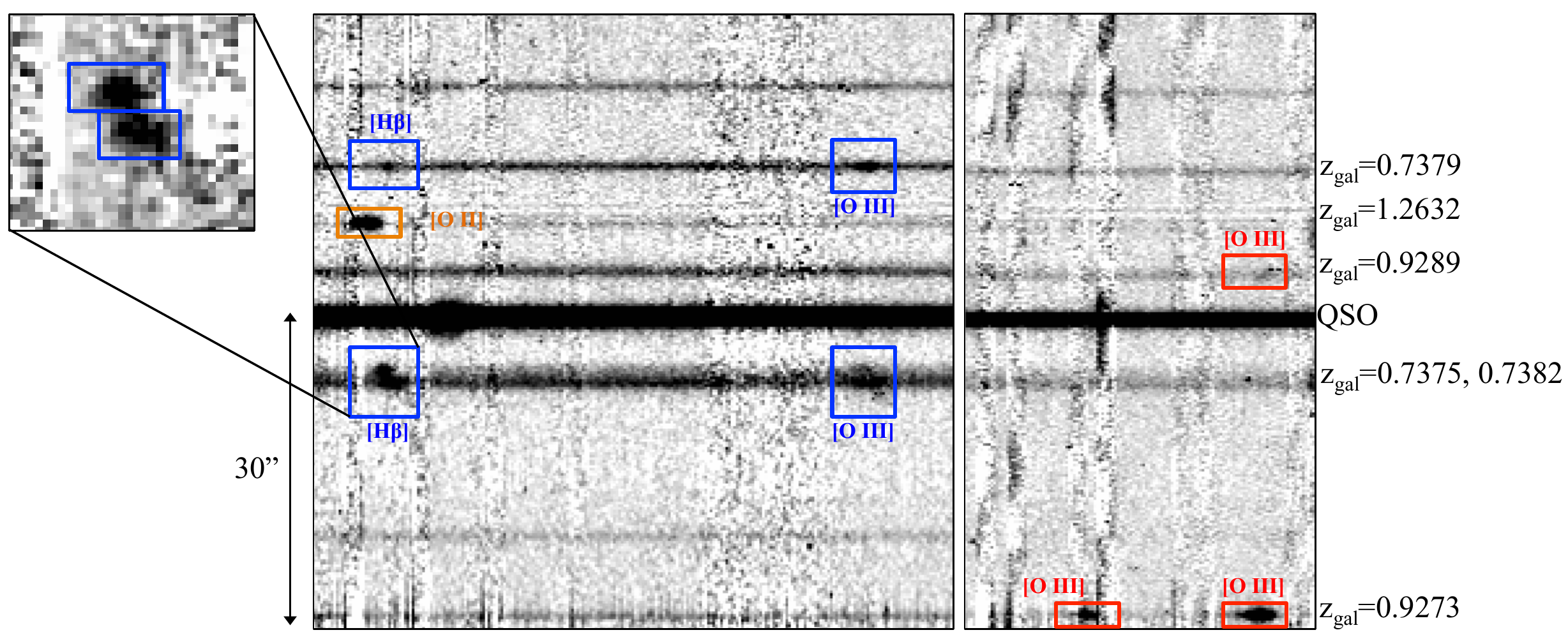}
    \caption{Colour-inverted FORS2 spectral segments of the Q\,1436-0051 field, observed during our campaign. Boxes highlight the identified emission lines, colour-coded to emphasize systems at similar redshift.}
    \label{fig:Q1436-0051_groupassociations}
\end{figure*}

%--------------------------------------------------------------------

\section{Notes in individual abundance measurements}
\label{sec:notes}

Here, we present the abundance measurements based on flux-ratios for individual objects. We evaluate the flux-ratios as the median-value with upper and lower uncertainties corresponding to the the 16th- and 84th percentile of the CDF. The CDF is generated from 1E6 realisations of the flux-ratios assuming normal-distributed fluxes centred on the measurement, with a width set to the associated flux uncertainty. In order to differentiate between abundances measured with different diagnostics, we use the following notation: $[\mathrm{O/H}]_\mathrm{diagnostic}$ where the subscript refers to the diagnostic line-ratio as defined in Equation \ref{eq:diagnostics} used. If the line-ratio is consistent with two abundances, we refer to these with `$l$' or `$u$' superscripts to denote the lower branch and upper branch value, respectively. When a comparison to the mass-metallicity relation is made, we refer to this value as $[\mathrm{O/H}]_\mathrm{MZ,M08}$. In this comparison, we do not consider internal scatter present in the M08 MZ-relation.

\begin{itemize}
    \item \textbf{HE\,1122-1649}: The $\mathrm{O}_2$ line-ratio gives an oxygen abundance $[\mathrm{O/H}]_{\mathrm{O}_2} = 8.70^{+0.17}_{-0.20}$. This is consistent with the value inferred from the mass-metallicity relation, $\mathrm{[O/H]}_{\mathrm{MZ,M08}} = 8.57$.
\newline
        
    \item \textbf{Q\,0153+0009}: The $\mathrm{O}_3$ line-ratio gives a double-valued abundance, with a lower- and upper  abundance branch value of $[\mathrm{O/H}]_{\mathrm{O}_3}^l = 7.02^{+0.03}_{-0.03}$ and $[\mathrm{O/H}]_{\mathrm{O}_3}^u = 8.75^{+0.03}_{-0.03}$, respectively. The upper-branch abundance is consistent with the value inferred from the mass-metallicity relation, $\mathrm{[O/H]}_{\mathrm{MZ,M08}} = 8.81$. Note that the reported uncertainties are underestimated. This is in part because we have not accounted for internal scatter in the M08 calibration, and in part because the abundance calibration is insensitive to variations in the flux-ratio around the measured value.
\newline
       
    \item \textbf{Q\,1209+107 PA2}: The $\mathrm{R}_{23}$ line-ratio, to within the $1\sigma$ uncertainty, does not constrain the abundance. Interpreting the large line-ratio as indicative of a unique abundance at the maximum line-ratio gives $\mathrm{[O/H]}_{\mathrm{R}_{23}} = 8.05$. The $\mathrm{O}_2$ line-ratio gives a double-valued abundance, with a lower- and upper branch value of $[\mathrm{O/H}]_{\mathrm{O}_2}^l = 8.55^{+0.05}_{-0.04}$ and $[\mathrm{O/H}]_{\mathrm{O}_2}^u = 8.83^{+0.03}_{-0.04}$, respectively. Likewise, the $\mathrm{O}_3$ line-ratio also gives a double-valued abundance, with a lower- and upper branch value of $[\mathrm{O/H}]_{\mathrm{O}_3}^l = 7.62^{+0.05}_{-0.04}$ and $[\mathrm{O/H}]_{\mathrm{O}_3}^u = 8.16^{+0.04}_{-0.05}$, respectively, and the $\mathrm{O}_{32}$ line-ratio gives an abundance $\mathrm{[O/H]}_{\mathrm{O}_{32}} = 8.35^{+0.02}_{-0.02}$. The inferred abundance from the mass-metallicity relation gives $\mathrm{[O/H]}_{\mathrm{MZ,M08}} = 8.02$. This is consistent with our interpretation of the $\mathrm{R}_{23}$. We therefore report the $\mathrm{R}_{23}$ value.
\newline
    
    \item \textbf{Q\,1436-0051}: We infer $f_{\mathrm{[\textsc{O iii}]} \lambda 4959}$ assuming the standard line-strength conversion $f_{\mathrm{[\textsc{O iii}]} \lambda 4959} = 0.34 f_{\mathrm{[\textsc{O iii}]} \lambda 5007}$. Combining our FORS2 measurements with our re-measured $f_{\mathrm{[\textsc{O ii}]}3727,3729}$ line-flux in the Magellan II spectrum \citep[][see Section \ref{sec:starformation}]{Straka2016}, we find an $\mathrm{R}_{23}$ line-ratio consistent with a double-valued abundance. The lower- and upper branch values are $\mathrm{[O/H]}_{\mathrm{R}_{23}}^l = 7.93^{+0.12}_{-0.33}$ and $\mathrm{[O/H]}_{\mathrm{R}_{23}}^u = 8.16^{+0.35}_{-0.11}$, respectively. The $\mathrm{O}_2$ line-ratio is outside the range to constrain the abundance. The $\mathrm{O}_3$ line-ratio gives an upper branch abundance $\mathrm{[O/H]}_{\mathrm{O}_3}^u = 8.82^{+0.08}_{-0.08}$, but we note that the flux-ratio is marginally consistent with a lower-branch abundance $\mathrm{[O/H]}_{\mathrm{O}_3}^l \sim 7.02$. The $\mathrm{O}_{32}$ line-ratio gives an abundance $\mathrm{[O/H]}_{\mathrm{O}_{32}} = 9.11^{+0.07}_{-0.06}$. We believe that the large scatter in abundances as determined with the different diagnostics is driven by the combination of spectroscopic data at different slit-placements and taken in different conditions. To avoid introducing such systematic differences in the abundance measurements, and noting the consistency with the value inferred from the mass-metallicity relation $\mathrm{[O/H]}_{\mathrm{MZ,M08}} = 8.92$, we assume the $\mathrm{[O/H]}_{\mathrm{O}_3}$ upper branch abundance.
\newline
        
    \item \textbf{Q\,2335+1501}: The $\mathrm{O}_2$ line-ratio gives a double-valued abundance, with a lower- and upper branch value of $[\mathrm{O/H}]_{\mathrm{O}_2}^l = 8.34^{+0.36}_{-0.19}$ and $[\mathrm{O/H}]_{\mathrm{O}_2}^u = 8.96^{+0.09}_{-0.26}$, respectively. We note that the upper branch abundance is consistent with the value inferred from the mass-metallicity relation, $\mathrm{[O/H]}_{\mathrm{MZ,M08}} = 8.82$.
\end{itemize}

\end{appendix}

\end{document}